\newcommand{\doublespace}{
  \renewcommand{\baselinestretch}{2.0}
  \large\normalsize}

\documentstyle[preprint,aps,tighten,prb]{revtex}
\begin{document}
\doublespace
\large

\centerline{\bf \Large 4. EFFECTS OF MICROSTRIAN ON BAND AND OPTICAL 
   PROPERTIES OF SILO QUANTUM WIRES} 




\section{Introduction}
\mbox{}

Optical properties of III-V semiconductor nanostructures have attracted 
a great deal of interest recently for their applications in optical communications that 
involve switching, amplification, and signal processing.


The fabrication of QWR's via the strain-induced lateral-layer ordering
(SILO) process starts with the growth of
short-period superlattices (SPS) [e.g. (GaAs)$_2$/(InAs)$_{2.25}$]
along the [001] direction. The excess fractional InAs layer leads to
stripe-like islands during the MBE growth.[4] The presence of stripes combined with the strain effect lead to a 
natural phase separation as
additional layers of GaAs or InAs are deposited.
A self-assembled QWR heterostructure can then be
created by sandwiching the composition modulated layer
between barrier materials such as Al$_{0.24}$Ga$_{0.24}$In$_{0.52}$As (quarternary), Al$_{0.48}$In$_{0.52}$As(ternary), or InP (binary)[4-6].
It was found that different barrier materials can lead to different degree of
lateral composition modulation, and the consequent optical properties are different.[6] Besides the self-assembled lateral
ordering, it is believed that the strain also plays a key role[5,6] 
in the temperature stability and optical anisotropy for 
the QWR laser structure.


	Our  theoretical study of Ga$_x$In$_{1-x}$As self-assembled QWRs in
Chapter 3 is
based on a simplified uniform strain model as used in few other studies[8-10].
In these calculations,
 the SPS region is modeled by a Ga$_x$In$_{1-x}$As alloy 
with a lateral modulation of the composition $x$. Although these calculations can explain
the QWR band gap and optical anisotropy qualitatively, it does not take into 
account the detailed SPS structure and the microscopic strain distribution.
The understanding of these effects is important if one wishes to have a full 
design capability of the self-assembled QWR optoelectronic devices. 
On the other hand, microscopic strain distributions in several SPS structures 
have been
calculated, and their electronic properties have been 
studied via an empirical pseudopotential method[11]. 
However, the effects of sandwiching the SPS structures between barrier 
materials so as to form QWRs  have not been
explored. It is worth pointing out that  
the effective masses obtained in the pseudopotential method for the 
constituent bulk materials are 0.032$m_0$ and 0.092$m_0$ 
for InAs and GaAs, respectively.[12] These values are  30\% larger
than the actual values (0.024$m_0$ and 0.067$m_0$); 
thus, the energy levels of quantum confined states obtained by this method 
are subject to similar uncertainty.  

In this chapter, we present a systematic study of the effects of microscopic 
strain distribution[11] on
the electronic and optical properties of the Ga$_{1-x}$In$_x$As 
self-assembled QWRs via the combination of
the effective bond-orbital model (EBOM) for electronic states and the 
valence-force-field (VFF) model for microscopic strain distribution.
Both clamped and unclamped SPS structures with different degrees of lateral 
alloy mixing are considered. The clamped structure
has a SPS layer sandwiched between the substrate and a thick capping layer
so that the interface between the top monolayer of the SPS and the barrier 
region is atomically flat.
The unclamped structures correspond to a SPS layer sandwiched between the 
substrate and a thin capping layer, which allows more flexible relaxation of
interface atoms so the strain energy can be relieved. 
This leads to a wavy interface structure. Most self-assembled QWR 
samples reported to date are closer to unclamped structures with a wavy 
interface, although it should be possible to produce self-assembled QWRs 
closer to the clamped structures by using a thick capping layer.

This study shows that there are profound differences in the 
electronic and optical properties between the clamped and unclamped self-assembled QWR 
structures. In particular, we find
that in clamped structures the electron and hole are confined in the Ga-rich 
region and the polarization of photoluminescence (PL) can go from 
predominantly along [110] for SPS with 
abrupt change in In/Ga composition to predominantly along [$1\bar 10$] for 
SPS with smooth change in In/Ga composition. On the other hand, for 
unclamped structures the 
electrons and holes are confined in the In-rich region, and the optical 
polarization is always predominately along [$1\bar 10$] with a 
weak dependence on the lateral alloy mixing. 
This implies that by changing the degree of strain relaxation at the 
interface between the  SPS and the capping layer, one can tailor the optical 
properties of self-assembled QWRs. Thus, through this study, 
we can gain a better understanding of the strain engineering of  
self-assembled QWR structures which may find applications in fiber-optical 
communication. 

\section{Model Structures}
\mbox{}

For both clamped and unclamped SPS structures, we consider varying degrees of
lateral alloy mixing and examine the effects of
the microscopic strain distribution on the electronic and optical properties.
Two example QWR model structures 
considered in the present paper (prior to alloy mixing) are depicted
in Figure 4.1. The supercell of the first model structure consists
of 8 pairs of (001) (GaAs)$_2$/(InAs)$_2$ SPS
with a total thickness of $\approx$ 94 \AA \,(quantum well region)
followed by a Al$_{0.24}$Ga$_{0.24}$In$_{0.52}$As layer (barrier region)
with thickness $\approx$ 60 \AA \,(20 diatomic layers).
The supercell of the second model structure consists
of 8 pairs of (001) (GaAs)$_2$/(InAs)$_{2.25}$ SPS
with a total thickness of $\approx$ 106 \AA \,(quantum well region)
followed by a Al$_{0.24}$Ga$_{0.24}$In$_{0.52}$As layer ( barrier region).
We assume an arrangement of alternating stripe like islands due to strain 
induced lateral ordering. In the diagram, no lateral alloy mixing is shown.
In our calculations of strain distribution and electronic structures, varying
degree of lateral alloy mixing will be considered.
Experimentally, self-assembled 
InGaAs QWRs were usually grown with the (2/2.25) SPS structure.[6] 
However, the migration
of excess In atoms during MBE growth could lead to
a structure somewhere between 
(2/2) SPS and (2/2.25) SPS structures.

In both model structures, we can divide the self-assembled QWR into
two regions with the left half being Ga rich and the right half In rich.
During growth, varying degree of lateral alloy
mixing of these islands with the surrounding atoms is likely to occur.
In the atomic layers with In-rich alloy filled the right half of the
unit cell (layer 2 in structure 1 and layers 7 and 9 in structure 2),  
the In composition $x_{\mbox{In}}$ is 
assumed to vary in the [110] direction (or $y'$ direction) 
according to the relation
\begin{equation}
x_{\mbox{In}}=  \left\{ \begin{array}{ll}
 x_m [1-\sin (\pi y'/2b)] /2 &  \mbox{ for } y'< b\\ 
	0  & \mbox{ for } b < y' < L/2-b \\
 x_m \{1 + \sin [\pi (y'-L/2)/2b]\} /2 &  \mbox{ for } L/2-b < y' < L/2+b \\
 x_m   &	  \mbox{ for } L/2+b < y' < L-b \\
 x_m \{1 - \sin [\pi (y'-L)/2b]\} /2 &  \mbox{ for } y'> L-b, \end{array} \right.
\end{equation}
where $x_m$ is the maximum In composition in the layer, $2b$ denotes the 
width of lateral composition grading, and $L$ is the period of the
lateral modulation in the [110] direction.
In layer 4 of structure 1, which contains Ga-rich alloy in the left half of the
unit cell, the Ga composition ($x_{\mbox{Ga}}$) as a function of $y'$ is given by
a similar equation with the sign of the sine function reversed as compared to
Eq. 4.1. In structure 2, there are a few atomic layers
that contain 0.25 monolayer of In (layers 3 and 13) or Ga (layers 5 and 11).
The lateral alloy modulation in layers 3 and 13 is described by
\begin{equation}
x_{\mbox{In}}=  \left\{ \begin{array}{ll}
 0   &    \mbox{ for } 0 < y' < 5L/8-b \\
 x_m \{1 + \sin [\pi (y'-5L/8)/2b]\} /2 &  \mbox{ for } 5L/8-b<y'< 5L/8+b,\\ 
 	x_m  & \mbox{ for } 5L/8+b < y' < 7L/8-b \\
 x_m \{1 - \sin [\pi (y'-7L/8)/2b]\} /2 &  \mbox{ for } 7L/8-b < y' < 7L/8+b \\
 0   &    \mbox{ for } 7L/8+b < y' < L.
\end{array} \right.
\end{equation}
Similar equation for $x_{\mbox{Ga}}$ in layers 5 and 11 can be deduced
from the above.
By varying the parameters $x_m$ and $b$, we can get different degrees
of lateral alloy mixing. Typically $x_m$ is between 0.6 and 1, and $b$
is between zero and $15 a_{[110]} \approx$ 62 \AA.

A VFF model[13-15] is used to find the equilibrium 
atomic positions in the self-assembled QWR structure by minimizing the lattice energy.
The strain tensor at each atomic (In or Ga) site is then obtained by
calculating the local distortion of chemical bonds.
We find that different local arrangement of atoms can lead to very different
strain distribution. In particular, the shear strain in the clamped structure
can change substantially when the In/Ga composition modulation is changed.
Consequently, the optical anisotropy can be reversed due to the change in 
the strength of the shear strain caused by
the intermixing of In and Ga atoms.

\section{Theoretical Approach}
\mbox{}

The method used in this paper for calculating the strained QWR band 
structure is based 
on EBOM[5] as described in Chapter 2. 
The model can be viewed as a spatially
discretized version of the ${\bf k\cdot p}$ method, while retaining the 
virtues of LCAO (linear combination of atomic orbitals) method.
The ${\bf k\cdot p}$ model is the most popular one for treating electronic
structures of semiconductor quantum wells or superlattices.
However, when applied to complex structures such as self-assembled 
quantum wires[4-9] or quantum dots[13,18,19], the method becomes very cumbersome
if one wishes to implement the correct boundary conditions that
take into account the differences in
${\bf k\cdot p}$ band parameters for different materials involved.
EBOM is free of this problem, since different material parameters are
used at different atomic sites in a natural way. For simple structures, when
both EBOM and ${\bf k\cdot p}$ model are equally applicable,
the results obtained are essentially identical.[16]

The optical matrix elements for the QWR states are computed
in terms of elementary optical matrix elements between the valence-band
bond orbitals and the conduction-band orbitals. 
The present calculation includes the coupling of the top four 
valence bands and the lowest two conduction bands (including spin degeneracy).
Thus, it is equivalent to a 6-band ${\bf k\cdot p}$ model.
For our systems studied here, the band-edge properties are relatively 
unaffected by the split-off band due to the large spin-orbit splitting
as discussed in our previous paper[9]. Hence, the split-off bands are 
ignored here. The bond-orbitals for the GaAs and InAs	needed in
the expansion of the superlattice states contain the following: 
four valence-band bond orbitals per bulk unit cell, which are p-like 
orbitals coupled with the spin to form orbitals with total 
angular momentum J=3/2 plus the products of the s-like conduction-band bond 
orbital and the electron spinors. 
The valence-band bond-orbitals are written as
\begin{equation}
|{\bf R},u_{JM}>=\sum_{\alpha,\sigma}C(\alpha,\sigma,J,M)|{\bf R},\alpha>
\chi_{\sigma},
\end{equation}
where $J= 3/2$, $M=-3/2,\cdots 3/2$, $\chi_{\sigma}$ designates the electron 
spinor ($\sigma$=1/2,-1/2), and $|{\bf R},\alpha>$ denotes an 
$\alpha$-like ($\alpha=x,y,z$) bond orbital located at unit cell 
${\bf R}$. $C(\alpha,\sigma,J,M)$ are the coupling coefficients 
obtainable by group theory. All these bond orbitals are assumed 
to be sufficiently localized so that the interaction between orbitals 
separated farther than the nearest-neighbor distance can be ignored. 

    The effect of strain is included by adding a strain Hamiltonian 
$H^{st}$ to the EBOM Hamiltonian[17].  The matrix elements 
of $H^{st}$ in the bond-orbital basis can be 
obtained by the deformation-potential theory of Bir and Pikus[20]. 
We use the VFF model of Keating[14] and Martin[15]
to calculate the microscopic strain distribution as described
 in Chapter 2. 
%
%
%
     
  To find the strain tensor in the InAs/GaAs self-assembled QWR, we start from  
ideal atomic positions and minimize the system energy 
using the 
Hamiltonian given above. Minimization of the total energy requires one to 
solve a set of coupled equations with $3N$ variables, where $N$ is the total 
number of atoms. Direct solution of these equations is impractical 
in our case, since the system contains more than 6,000 atoms. 
We use an approach taken by several authors[10,13,19] which has been shown 
to be quite efficient. In the beginning of the simulation all the atoms are 
placed on the InP lattice, we allow atoms to deviate from this starting 
positions and use periodic boundary conditions in the plane
perpendicular to the growth direction, while keeping atoms in the
planes outside the SPS region at their ideal atomic positions
for an InP lattice (since the self-assembled QWR is grown epitaxially on the InP 
substrate). In each iteration, only one atomic 
position is displaced and other atomic positions are held fixed. 
The direction of the displacement of atom $i$ is determined according to
the force  $f_i=-\partial V/\partial x_i$ acting on it.
All atoms are displaced in sequence. The whole sequence is
repeated until the forces acting on all atoms become zero,
at which point the system energy is a minimum.
Once the positions of all the atoms are known, the strain distribution 
is obtained through the strain tensor calculated according to the method
described in Equation 2.5 [21] as described in Chapter 2.
%
%
%
%

Once the strain tensor is obtained, the strain Hamiltonian is given by Bir and
Pikus[20]

\begin{equation}
H^{st} = \left( \begin{array}{ccc} 
 -\Delta V_H + D_1 &\sqrt{3}d e_{xy} &\sqrt{3}d e_{xz} \\
  \sqrt{3}d e_{xy} & -\Delta V_H + D_2     &\sqrt{3}d e_{yz} \\ 
       \sqrt{3}d e_{xz} &\sqrt{3}d e_{yz} & -\Delta V_H + D_3
               \end{array} \right),
\end{equation}
where $e_{ij}=(\epsilon_{ij}+\epsilon_{ji})/2$, and
$$
\Delta V_H = (a_1+a_2)(\epsilon_{xx}+\epsilon_{yy}+\epsilon_{zz}),~~\\
 D_1 = b(2\epsilon_{xx}-\epsilon_{yy}-\epsilon_{zz}),~~\\$$
$$
 D_2 = b(2\epsilon_{yy}-\epsilon_{xx}-\epsilon_{zz}),~~\\
 D_3 = b(2\epsilon_{zz}-\epsilon_{yy}-\epsilon_{xx}).\\
$$
The strain potential on the $s$ states is given by
$$\Delta V_c = c_1(\epsilon_{xx}+\epsilon_{yy}+\epsilon_{zz}),$$
The strain Hamiltonian in the bond-orbital basis $|JM>$ can be easily 
found by using the coupling constants[13], i.e, 
\begin{equation}
<JM|H^{st}|J'M'>=\sum_{\alpha, \alpha',\sigma}C(\alpha, \sigma;J,M)^*
C(\alpha', \sigma;J',M')H^{st}_{\alpha\alpha'}
\end{equation}
The elastic constants $C_{12}$ and $C_{11}$ for GaAs, InAs and AlAs can be found
in Refs. [22,23]. The deformation potentials $a_1,~a_2,~b,~c_1, ~d$ can be
found in Refs. [23-26]. The linear interpolation
and virtual crystal approximations are used to obtain the corresponding
parameters for the barrier material (Al$_{0.24}$Ga$_{0.24}$In$_{0.52}$As).

The above strain Hamiltonian is derived locally for the each cation atom in 
the self-assembled QWR considered. To calculate the electronic states of the 
self-assembled QWR for model structure 1 [Fig.4.1(a)],
we first construct a zero$th$-order
Hamiltonian for a superlattice structure which contains in each period
8 pairs of (001) (GaAs)$_{2}$/(InAs)$_{2}$ 
SPS layers (with a total thickness around 94 \AA) and 20 diatomic layers of 
Al$_{0.24}$Ga$_{0.24}$In$_{0.52}$As (with thickness around 60 \AA).
So, the superlattice unit cell for the zero$th$-order model contains
52 diatomic layers.  The appropriate strain Hamiltonian for the
the (GaAs)$_{2}$/(InAs)$_{2}$ SPS  on InP is also included.
For model structure 2 [Fig. 4.1(b)], the zero$th$-order superlattice
consists of two additional monolayers of InAs 
inserted into the 8 pairs of (2/2) SPS:
one between the 2nd and 3rd pair of (2/2) SPS, the other between the 
5th and 6th pair of (2/2) SPS. 

The eigen-states for the zero$th$-order Hamiltonian for different
values of $k_2$ (separated by the SL reciprocal lattice vectors
in the [110] direction) are then used as the basis
for calculating the self-assembled QWR electronic states.
The difference in the Hamiltonian (including strain effects)
caused by the intermixing of Ga and In
atoms at the interfaces is then added to the zero$th$-order Hamiltonian,
and the electronic states of the  full Hamiltonian is solved by 
diagonalizing the Hamiltonian matrix defined within a truncated set of
eigen-states of the zero$th$-order Hamiltonian. A total of $\sim 500$ eigen-states
of the zero$th$-order Hamiltonian (with 21 different $k_{110}$ points)
were used in the expansion. The subbands
closest to the band edge are converged to within 0.1 meV.

\section{Results and Discussions}

{\bf A. Strain distributions}

In this section we discuss strain distributions in our model structures
as described in Sec. 4.2. Both clamped and unclamped 
situations are considered.
To model the alloy structure with composition modulation,
we use a super-cell
which contains 72 atoms in the [110] ($y'$) direction, 36 atoms in the
[$1\bar 1 0$] ($x'$) direction and 64 or 68 atomic planes along the [001] ($z$)
direction [8 pairs of (2/2) or (2/2.25) SPS] plus a GaAs capping layer. 
In the atomic planes which consist of alloy structure, we first determine
the In composition at a given $y'$ according to either Eq. 4.1 or 4.2 and 
then use a random number generator to determine the atomic species along the
$x'$ direction. The bottom layer of atoms are bonded to the InP substrate
with the substrate atoms fixed at their ideal atomic positions.
The calculated strain distributions are then averaged over
the $x'$ coordinate. 
For the clamped case,  the GaAs capping layer is assumed to be lattice
matched to InP with a flat surface. For the unclamped case, the capping
layer is allowed to relax freely, thus giving rise to a wavy surface structure.

The diagonal strains in
the four atomic layers that constitute the (2/2) SPS in structure 1 for 
the clamped and unclamped cases are shown in Figs. 4.2 and 4.3, respectively.
The lateral alloy modulation considered is described by Eq. 4.1
with $x_m=0.8$ and $b=7a_{[110]}$.
For best illustration, we show diagonal strains  
in a rotated frame, in which $x'$ is [1$\bar 1$0], $y'$ is [110], 
and  $z'$ is [001].  The layer number in the figure labels the atomic
layers in Fig. 4.1(a), starting from the bottom layer.
There are two main features worth pointing out in Figs. 4.2 and 4.3. First, 
in the ideal situation (without atomic relaxation), one would predict 
$\epsilon_{y'y'}$ to be the same as $\epsilon_{x'x'}$ due to symmetry. 
However, with atomic relaxation, all Ga(In) atoms tend to shift
in a direction so as to reduce the strain in the Ga(In)-rich region.
Thus, the magnitude of $\epsilon_{y'y'}$ on Ga(In) sites
(dashed lines) is lower (higher) than $\epsilon_{x'x'}$ in
Ga(In)-rich region.
Second, the $z'$ component strain (dash-dotted lines) tends to 
compensate the other two components such that the volume of each bulk unit cell
is closer to that for the unstrained bulk.
Thus, we see that $\epsilon_{z'z'}$  has an opposite sign
compared to $\epsilon_{x'x'}$ or $\epsilon_{y'y'}$ at all atomic sites. 

Comparing Fig. 4.2 with Fig. 4.3, we see that the main difference between
the clamped and unclamped case is that in the atomic layers with
lateral composition modulation (layers 2 and 4), the hydrostatic strain
(sum of all three diagonal strain components) is much smaller in the 
unclamped structure than in the clamped structure. This would lead
to a major difference in the band-edge profile for the two structures
to be discussed below.

	The difference in $\epsilon_{x'x'}$ and $\epsilon_{y'y'}$ 
leads to a nonzero shear strain $e_{xy}=(\epsilon_{y'y'}-\epsilon_{x'x'})/2$
in the original coordinates. For the clamped case, 
the shear strain is found to be particularly strong near the boundary where
the alloy composition begins to change, 
and it is sensitive to the degree of lateral alloy mixing.
For $x_m=0.8$ and $b=7a_{[110]}$, the maximum value of
$\epsilon_{xy}$ is around 0.4\%. 
For abrupt composition modulation ($x_m=1$ and $b=0$)
the  maximum $\epsilon_{xy}$ value increases five-fold to around 2\%.
The other shear strain components ($\epsilon_{xz}$ and
$\epsilon_{yz}$) are found to have similar magnitude.
For the unclamped case, the shear strain is strong even in regions
away from the boundary, and the maximum shear strain is larger than 
its counterpart in the clamped structure by about 30\%.

The diagonal strains of structure 2 for the unclamped case 
with $x_m=1$ and $b=7a_{[110]}$
are shown in Fig. 4.4. The layer number in the figure labels the atomic
layers in Fig. 4.1(b), starting from the bottom layer.
Only four representative atomic layers (3,4,5, and 7)
are shown. We note that the average magnitude of the hydrostatic strain in the 
In-rich region in model structure 2 is comparable to that for 
unclamped structure 1.

\noindent {\bf B. Electronic and optical properties}

    In order to understand the aspect of lateral quantum confinement
due to composition modulation and strain, we examine
the band-edge energies of a strained quantum well structure whose
well material is the same as appeared in the self-assembled
QWR with a fixed value of $y'$ and the barrier consists of 60 \AA \, thick
Al$_{0.24}$Ga$_{0.24}$In$_{0.52}$As.
For structure 1 depicted in
Fig. 4.1(a), the well material consists of 8 pairs of
(GaAs)$_2$/(InAs)$_2$ SPS.
The conduction band minimum and valence band maximum of the above quantum well 
with varying degree of lateral alloy mixing
as functions of $y'$ are shown in Fig. 4.5 for both clamped
and unclamped cases. 
The strain Hamiltonian used here is the same as that used in
the self-assembled QWR at the corresponding $y'$.
All material parameters are chosen to be
the same as in Ref. [9] at 77K, except that the deformation
potential $c_1$ of GaAs is slightly modified from $-6.8$ eV to $-7.1$ eV so
that $a_1+a_2+c_1=-9.8$ eV is in agreement with the experimental measurements[26]
and the valence-band offset between GaAs and InAs used here is 0.26 eV 
according to the model solid theory[27].
To correct the band gap of SPS and InGaAs alloy due to the bowing effect, we
add a correction term $-0.4x(1-x)$ eV to the diagonal element for the
$s$-like bond orbital in the Hamiltonian, where $x$ is the effective alloy
composition of the SPS or alloy at a given $y'$.
With the correction of bowing effect, our model gives band gaps for
the (GaAs)$_2$(InAs)$_2$ SPS, Ga$_{1-x}$In$_{x}$As alloy, and 
the superlattice made of (GaAs)$_2$(InAs)$_2$ SPS grown on InP substrate
all in very good agreement with available experimental observations.
The PL measurements
indicate that the (GaAs)$_2$(InAs)$_2$ SPS grown on InP
substrate has a band gap around 0.75 eV.[28] Here we obtain a band gap of
0.74 eV for the (GaAs)$_2$(InAs)$_2$ SPS with a strain distribution 
obtained again via the VFF model and
0.79 eV for 8 pairs or 100 \AA \, of (GaAs)$_2$(InAs)$_2$ SPS 
sandwiched between Al$_{0.24}$Ga$_{0.24}$In$_{0.52}$As confining barriers.
This is also consistent with the PL measurements on the
(GaAs)$_2$(InAs)$_2$/InP multiple quantum wells.[28] 

The band-edge profiles shown in Fig. 4.5 suggest that for the clamped 
case with alloy mixing, both electrons
and holes are confined in the Ga-rich region with an effective barrier height
around $0.13$ eV for the electrons and $0.1$ eV for the holes. 
Both offsets are large enough
to give rise to strong lateral confinement for electrons and holes in
the Ga-rich region. Without alloy mixing, the electrons are not well confined,
since the average energy between the Ga-rich and In-rich regions are nearly
the same. For the unclamped case, both electrons
and holes are confined in the In-rich region with an effective barrier height
around $0.2 - 0.3 eV$ for the electrons and 0.1 - 0.2 eV for the holes. 
We note that the band gap in the In-rich region is rather insensitive to
the alloy mixing, while the effective barrier height can change substantially
due to different alloy mixing. This is because the band gap changes in In-rich
region due to the alloy mixing alone and due to the change in strain 
distribution happen to nearly cancel each other for the degree of 
mixing considered here.

Figure 4.6 shows the band-edge profiles for structure 2 
[8 pairs of (2/2.25) SPS sandwiched between 60\AA \,
Al$_{0.24}$Ga$_{0.24}$In$_{0.52}$As barriers]
with and without alloy mixing for both clamped and unclamped cases.
For the clamped case, the electrons are confined in the
In-rich region, while the holes are confined in the Ga-rich region. Thus, we 
have a spatially indirect QWR, which will have very weak interband optical 
transition strength.  For the unclamped case, the band-edge profile is similar to
the unclamped case of structure 1 with both the electrons and holes confined in
the In-rich region. The band gaps for all cases of structure 2 are consistently
smaller than their counterparts in structure 1.    

Figure 4.7 shows the near zone-center valence subband structures and squared 
optical matrix elements ($P^2$) for the QWR model structure 1 with clamped 
case and without considering alloy mixing.
Here $P^2$ is defined
as $\frac 2 {m_0} 
\sum_{s,s'} |<\psi_{v}|\hat \epsilon \cdot {\bf p}|\psi_{c}>|^2$,
where $\psi_{v}$ ($\psi_{c}$) denotes a valance (conduction) subband state.
The symbol $\sum_{s,s'}$ means a sum over two nearly degenerate pairs of
subbands in the initial and final states. $\hat \epsilon$ denotes
the polarization vector of light. We consider only the $x'$ (along the
QWR axis) and $y'$ (perpendicular to the QWR axis) components of the
polarization vector.
Only the dispersion along the $k_1$ ($[1\bar 10]$) 
direction is shown in Fig. 4.7, since
the dispersion along the $k_2$ direction for the confined levels  
is rather small due to strong lateral confinement.
All subbands are two-fold degenerate at the zone center
due to the Kramer's degeneracy and
they split at finite wave vectors as a result of lack of
inversion symmetry in the system.
The first three pairs of subbands are labeled V1, V2, and V3.
They have unusually large energy separations compared with other
valence subbands. This is because the first three pairs of subbands represent
QWR confined states, whereas the other valence subbands 
are unconfined in the $y'$ direction.

To examine the effects caused by the shear strain, we have also calculated
the band structures with the shear strain set to zero. We found that
without the shear strain the first pairs of valence subbands (V1) have 
much larger dispersion along the $k_1$ direction with an effective mass
along $[1\bar 10]$ about a factor five smaller than that  
with the shear strain. This indicates that the
V1 subband states for the case without shear strain are derived
mainly from bond orbitals with $x'$-character, which leads to stronger
overlap between two bond orbitals along the $x'$ direction, hence 
larger dispersion along $k_1$. When the shear strain is present, the character
of bond orbitals in the V1 subband states change from $x'$-like 
to predominantly $y'$-like; thus, the dispersion along $k_1$ becomes
much weaker. This explains the very flat V1 subbands as shown in Fig. 4.7.
The switching of orbital character in the V1 subbands can be
understood as follows. When the  shear strain is absent, we have 
$\epsilon_{x'x'}=\epsilon_{y'y'}$, and the confinement effect in the $y'$
direction pushes down the states with $y'$ character (which has smaller
effective mass in the $y'$ direction), thus leaving the top valence 
subband (V1) to have predominantly $x'$ character. On the other hand,
with the presence of shear strain as shown in structure 1, 
we have $\epsilon_{y'y'} < \epsilon_{x'x'}$ in Ga-rich region, 
and the bi-axial term of the strain potential forces the $y'$-like states 
to move above the $x'$-like states, overcoming the confinement effect. 
As a result, the V1 subbands become $y'$-like.
The conduction subbands are approximately
parabolic as usual with a zone-center subband minimum equal to 688 meV. 
This gives an energy gap of 763 meV  for QWR model structure 1.
As seen in Fig. 4.7, the C1-V1 transition strength for $y'$-polarization (dashed
line) is more than twice that for the $x'$-polarization (solid line) 
[$P^2_{[1\bar 1  0]}/P^2_{[110]} \approx 0.36$].
This is consistent with the fact that the bond orbitals involved in the
V1 subbands are predominantly $y'$-like.

	We have also studied the case with 
alloy mixing for clamped structure 1 in which a gradual lateral modulation in In composition 
substantially reduces the shear strain. Using a lateral alloy mixing
described in Eq. 4.1 with $x_m=0.8$ and $b=7a_{[110]}$,
we find significant change in band structures and optical properties
compared to the case without alloy mixing.
With alloy mixing we found that the first pairs of valence subbands 
(V1) have  much larger dispersion than their counterparts in Fig. 4.7, and
the  atomic character in the $V1$ states changes from predominantly
$y'$-like to $x'$-like.  
The change of atomic character in the V1 states
is a consequence of the competition between the QWR confinement which
suppresses
the $y'$ character and the $e_{xy}$ shear strain which suppresses the $x'$
character. As a result, the ratio $P^2_{[1\bar 1  0]}/P^2_{[110]}$ changes 
from 0.36  to 3.5, indicating a reversal of optical anisotropy.

Next, we study the properties of unclamped self-assembled QWRs. 
We find that for this case, the electrons and holes are always 
confined in the In-rich region, regardless of the degree of alloy mixing 
and whether the QWR structure consists of (2/2) SPS or (2/2.25) SPS. 
Furthermore, since the strain in the In-rich region has opposite sign compared to
that in the Ga-rich region, the shear strain has an opposite effect
on the VB states confined in the In-rich region versus the Ga-rich region. 
Namely, it pushes the energy of the
$y'$-like VB states down relative to the $x'$-like VB states. 
Thus, both the QWR
confinement and the shear strain effects are in favor of giving a predominantly 
$x'$ character in the V1 subband. Therefore, the interband optical 
transition always has a ratio $P^2_{[1\bar 1  0]}/P^2_{[110]}$ larger than 1.
  
To illustrate this, we show in Fig. 4.8 the near zone-center valence subband 
structures and squared optical matrix elements ($P^2$) for
the QWR model structure 1 as depicted in Fig. 4.1(a) with thin capping
layer and with alloy mixing described by $x_m=0.8$ and
$b=7a_{[110]}$. 
Comparing the band structure in Fig. 4.8 with that in Fig. 4.7, we
see that the V1 subband has much larger dispersion (i.e. much smaller
effective mass) than its counterpart in Fig. 4.7. There are two reasons for this.
First, the holes in the unclamped case are now confined in the In-rich region
rather than in the Ga-rich region as in the clamped case, thus having
much smaller effective mass. Second, both the QWR confinement
and the shear strain effects lead to a predominantly $x'$ character in the V1 
states, which tend to yield a smaller effective mass in the $k_1$ direction.
The optical properties shown in Fig. 4.8 are also distinctly different from that 
shown in Fig. 4.7. The interband optical transition (C1-V1) now shows a
much stronger polarization along [1$\bar 10$] than [110]. 
The quick drop of the the $y'$ polarization in C1-V1 transition
at $k_1\approx 0.035 \frac {2\pi} a$ is caused by the band
mixing of V1 and V2 states
as can be seen in the VB band structures in this figure.

We have also calculated the near zone-center valence subband structures 
and squared optical matrix elements ($P^2$) for
the QWR model structure 2 with thin capping
layer (not shown), and we find that the band structures and optical matrix 
elements are similar to Fig. 4.8, except that the energy separation between 
V1 and V2 subbands is smaller (by $\sim$ 5 meV) and the polarization ratio 
($P^2_{[1\bar 1  0]}/P^2_{[110]}$) is larger. 

	Finally, we list in Table 4.1
the band gaps and the polarization ratio ($P^2_{[1\bar 1  0]}/P^2_{[110]}$) 
for various
QWR structures with varying degrees of lateral alloy mixing (as indicated by
the parameters $x_m$ and $b$) with the last line as the experimental results.
To calculate the polarization ratio in the PL spectra, we integrate the squared optical matrix
element over the range of $k_1$ corresponding to the spread of exciton 
envelope function in the $k$ space. The exciton envelope function
is obtained by solving the 1D Schr\"{o}dinger equation for the exciton in
the effective-mass approximation similar to what we did in Ref. 9.
This table allows one to see 
the trend of optical properties as the degree of lateral alloy mixing varies 
and as the structure changes from (2/2) SPS to (2/2.25) SPS. 
For clamped (2/2) SPS structure, the band gap is insensitive to alloy 
mixing, but the polarization ratio changes drastically due to the 
shear strain effect.  The clamped (2/2.25) SPS structure is spatially 
indirect, hence is not listed here.
For unclamped (2/2) SPS structure, the band gap varies between 0.77 and 0.8 eV and the polarization ratio changes from 3.1 to 1.5 as the degree of alloy mixing is increased. For unclamped (2/2.25) SPS structure, both the band gap (around 0.74 eV) and the 
polarization ratio (around 2) are insensitive to the change in degree of 
alloy mixing.

	Experimentally, the PL peak for self-assembled InGaAs QWRs
with similar dimensions as considered in our model structures
is around 735 meV and the polarization ratio for different samples is 
found to be between 2 and 4.[4-6] Taking into account an exciton binding 
energy between 10 and 20 meV, we find that our theoretical predictions 
for both unclamped and clamped structures with lateral alloy mixing (after we
included the bowing correction for InGaAs alloy and SPS) are 
consistent with the experimental findings. Since different samples can have
different structures, it requires more detailed information about the
electronic states (such as the energy levels of various electron and
hole subbands and the associated interband transition strengths) 
in order to determine the
actual geometry. Such information may be obtained via comparison 
between the theory and experiment for the interband or inter-subband
absorption spectra.  
 
\section{Conclusion}
\mbox{}

We have calculated the band structures and optical matrix elements 
for the self-assembled GaInAs  QWR grown by the SILO method. The actual SPS
structure and the microscopic strain distribution have been
taken into account. The effects of microscopic strain  distribution
on the valence subband structures and optical matrix elements are studied
for two types of self-assembled QWR structures, 
one with (2/2) SPS structure and the other with (2/2.25) SPS structure, 
with varying degrees of lateral alloy mixing. The clamping effect due to thick capping layer is also examined. 
The VFF model is used to calculate the equilibrium
atomic positions in the QWR model structures. This allows the calculation
of the strain distribution at the atomistic level.

We find that in model structures with thick capping layer (clamped case),
the magnitude of shear strain is quite large, around 2\%, when the lateral
alloy mixing is absent, and it reduces substantially when the lateral 
alloy mixing is introduced.  We found that the shear strain effect can 
alter the valence subband structures substantially when the holes are 
confined in the Ga-rich region, and it gives rise
to a reversed optical anisotropy (i.e. $P^2_{[1\bar 1  0]} < P^2_{[110]}$)
compared with QWR structures with negligible shear strain. 
This points to a possibility of "shear strain-engineering" to obtain
QWR laser structures of desired optical anisotropy.
For model structures with thin capping layer (unclamped case), the effect of
shear strain is less dramatic, since the holes are confined in the In-rich region,
where both the QWR confinement and the shear strain have similar effect. 
We found that polarization ratio ($P^2_{[1\bar 1  0]}/P^2_{[110]}$) 
for the unclamped case is always larger than 1 for both (2/2) and 
(2/2.25) SPS structures, regardless of the degree of lateral alloy mixing. 

    The band gap and polarization ratio obtained for both clamped and unclamped (2/2) SPS structures with lateral alloy mixing ($x_m \approx 0.8$) and for unclamped (2/2.25) SPS structures  are consistent
with experimental observations on most self-assembled InGaAs 
QWRs with similar specifications. For clamped (2/2) SPS structure without alloy mixing, we predicted a reversed optical anisotropy compared with the PL
measurements for most InAs/GaAs self-assembled QWRs. However, there exist a few
InAs/GaAs self-assembled QWR samples which display the reversed optical 
anisotropy at 77 K[7], and recent experimental studies also indicated a reversed
optical anisotropy in many samples at 300 K.[6] Our studies demonstrated 
a possibility for the reversed optical anisotropy that is due to the shear 
strain effect. 
There may be other possibilities for the reversed optical anisotropy at 300K.
For example, the thermal population of the excited hole states, which 
have different atomic characters from the lowest-lying hole states as 
may also lead to a change in the polarization ratio.
The temperature dependence of the PL peak position and polarization in 
self-assembled InGaAs QWRs[6] remains an intriguing question to be 
addressed in the future.

\begin{table}[h]
\caption{\bf \large List of band gaps and polarization ratios for 
all structures studied}

\vspace{1ex}
\begin{tabular}{lllllll} \hline\hline
$x_m$ \hspace{0.1in} & b/$a_{110}$\hspace{0.05in} & alloy mixing
\hspace{0.05in} & clamped/unclamped\hspace{0.05in} & SPS structure
\hspace{0.05in} & gap(eV)\hspace{0.1in} & $P^2_{\parallel}/P^2_{\perp}$ \\ \hline
 1.0  & 0   & no       & clamped       & (2/2)         & 0.765   & 0.36  \\ 
0.8   & 7  & yes       & clamped        & (2/2)         & 0.763   & 3.5  \\ 
 1.0  & 0   & no       & unclamped     & (2/2)         & 0.791   & 3.1  \\ 
 0.7  & 7  & yes       & unclamped      & (2/2)         & 0.766   & 1.8  \\ 
 0.6  & 7  & yes       & unclamped      & (2/2)         & 0.80    & 1.82 \\ 
0.6   & 15 & yes       & unclamped      & (2/2)         & 0.798   & 1.5  \\ 
1.0   & 0   & no       & unclamped     & (2/2.25)      & 0.731   & 2.2 \\
1.0   & 7  & yes       & unclamped      & (2/2.25)       & 0.74    & 2.1 \\ 
Expts[4] &~  &  ~& ~             &~               &0.735    &2-4 \\
\hline\hline
\end{tabular}

\end{table}

\newpage

\mbox{}
\mbox{}

\clearpage
\mbox{}
\vspace{3.0in}
\begin{figure}[h]
\caption{\bf \large  Schematic sketch of the unit cell of
the self-assembled quantum wire for two model structures considered. 
Each unit cell consists of 8 pairs of (2/2) or (2/2.25) GaAs/InAs 
short-period superlattices (SPS).
In structure 1, the (2/2) SPS contains four diatomic layers (only the cation
planes are illustrated) and the SPS is repeated eight times. 
In structure 2, four pairs of (2/2.25) SPS (or 17 diatomic layers) form
a period, and the period is repeated twice in the unit cell.  
Filled and open circles indicate Ga and In rows 
(each row extends infinitely along the $[1\bar 1 0]$ direction). }
\end{figure}


\mbox{}
\mbox{}

\clearpage
\begin{figure}[b]
\caption{ \bf \large Diagonal strain distribution of model structure 1
[corresponding to Fig. 4.1(a)] with thick capping layer (clamped case).
Solid: $\epsilon_{x'x'}$. Dashed: $\epsilon_{y'y'}$.
Dash-dotted: $\epsilon_{z'z'}$. Parameters for alloy mixing: $x_m=0.8$ and 
$b=7 a_{[110]}$.         }
\end{figure}

\mbox{}
\mbox{}
\clearpage
\begin{figure}[b]
\caption{\bf \large Diagonal strain distribution of model structure 1
with thin capping layer (unclamped case).
Solid: $\epsilon_{x'x'}$. Dashed: $\epsilon_{y'y'}$.
Dash-dotted: $\epsilon_{z'z'}$.
Parameters for alloy mixing: $x_m=0.8$ and $b=7 a_{[110]}$.}
\end{figure}

\mbox{}
\mbox{}

\clearpage
\begin{figure}[b]
\caption{ \bf \large Diagonal strain distribution of model structure 2
[corresponding to Fig. 4.1(b)] with thin capping layer (unclamped case).
Solid: $\epsilon_{x'x'}$. Dashed: $\epsilon_{y'y'}$.
Dash-dotted: $\epsilon_{z'z'}$.
Parameters for alloy mixing: $x_m=0.8$ and $b=7 a_{[110]}$.}         
\end{figure}

\mbox{}
\mbox{}

\clearpage
\mbox{}
\vspace{3.0in}
\begin{figure}[h]
\caption{\bf \large Conduction and valence band edges
for 8 pairs of (2/2) SPS structures
sandwiched between 60\AA Al$_{0.24}$Ga$_{0.24}$In$_{0.52}$As barriers
with different degrees of lateral 
alloy mixing as functions of the [110] coordinate $y'$ for both
clamped (thick capping layer) and unclamped (thin capping layer) case.
In upper panel, the dashed line is for abrupt modulation (without alloy mixing)
and the solid line is for lateral alloy mixing with $x_m=0.8$ and $b=7 a_{[110]}$.
In lower panel, dash-dotted: without alloy mixing; solid:
$x_m=0.7$ and $b=7 a_{[110]}$;
dotted: $x_m=0.6$ and $b=7 a_{[110]}$;
dashed: $x_m=0.6$ and $b=15 a_{[110]}$.}
\end{figure}

\mbox{}
\mbox{}

\clearpage
\begin{figure}[b]
\caption{\bf \large  Conduction and valence band edges
for 8 pairs of (2/2.25) SPS structures
sandwiched between 60\AA Al$_{0.24}$Ga$_{0.24}$In$_{0.52}$As barriers
with different degrees of lateral 
alloy mixing as functions of the [110] coordinate $y'$ for both
clamped (thick capping layer) and unclamped (thin capping layer) case.
Dashed: without alloy mixing.
Solid: with alloy mixing described by $x_m=1.0$ and $b=7 a_{[110]}$.}
\end{figure}

\mbox{}
\mbox{}

\clearpage
\begin{figure}[b]
\caption{\bf \large Valance subband structures and squared optical matrix elements
for self-assembled QWR made of (2/2) SPS structure                
with thick capping layer (clamped case) and without alloy mixing.      
}
\end{figure}

\mbox{}
\mbox{}

\clearpage
\begin{figure}[b]
\caption{\bf \large Valance subband structures and squared optical matrix elements
for self-assembled QWR made of (2/2) SPS structure              
with thin capping layer (unclamped case) and with alloy mixing described
by $x_m=0.8$ and $b=7 a_{[110]}$.      }
\end{figure}


\newpage

\begin{thebibliography}{wide label}

\bibitem{adams}A. R. Adams, Electron. Lett. {\bf 22}, 249 (1986).
\bibitem{yablonocitch}E. Yablonovitch and E. O. Kane. IEEE J. Lightwave 
Technol. LT-4, 504 (1986).
\bibitem{agrawa}G. P. Agrawa and N. K. Dutta, Long Wavelength Semiconductor
 Lasers, 2nd ed. (Van. Nostrand Reinhold, New York, 1993) Chap.7.
\bibitem{chou}S.T. Chou, K. Y. Cheng, L. J. Chou, and K. C. Hsieh, Appl. 
Phys. Lett. {\bf 17}, 2220 (1995); J. Appl. Phys. {\bf 78} 6270, (1995); 
J. Vac. Sci. Tech. B {\bf 13}, 650 (1995);
K. Y. Cheng, K. C. Hsien, and J. N. Baillargeon, Appl. Phys. Lett. {\bf 60}, 
2892 (1992).
\bibitem{wohlert}D. E. Wohlert, S. T. Chou, A. C Chen, K. Y. Cheng, 
and K. C. Hsieh, Appl. Phys. Lett. {\bf 17}, 2386 (1996).
\bibitem{wohlert1} D.E. Wohlert, and K. Y. Cheng, Appl. Phys. Lett. {\bf 76}, 2249 (2000).
\bibitem{wohlert2} D.E. Wohlert, and K. Y. Cheng, private communications. 
\bibitem{rich} Y. Tang, H. T. Lin, D. H. Rich, P. Colter, and S. M. Vernon,
Phys. Rev. B {\bf 53}, R10501 (1996).
\bibitem{masc} Y. Zhang and A. Mascarenhas, Phys. Rev. B{\bf 57}, 12245
(1998).
\bibitem{li}L. X. Li and Y. C. Chang, J. Appl. Phys. {\bf 84} 6162, 1998. 
\bibitem{Zun1} T. Mattila, L. Bellaiche, L. W. Wang, and A. Zunger, Appl.
Phys. Lett., {\bf 72}, 2144 (1998).
\bibitem{Zun2} L. W. Wang, J. Kim, and A. Zunger, Phys. Rev. B {\bf 59}, 5678
(1999).
\bibitem{jiang}H. Jiang and J. Singh, Phys. Rev. B{\bf 56}, 4696(1997).
\bibitem{keating}P. N. Keating, Phys. Rev. {\bf 145}, 637(1966).
\bibitem{martin}R. M. Martin, Phys. Rev. B{\bf 1}, 4005(1969).
\bibitem{chang}Y. C. Chang, Phys. Rev. B{\bf 37}, 8215 (1988).
\bibitem{houng}Mau-Phon Houng and Y.C Chang, J. Appl. Phys. {\bf 65}, 3096
(1989).
\bibitem {Gru} M. Grundmann, O.Stier, and D. Bimberg, 
Phys. Rev. B {\bf 52},11969 (1995).
\bibitem {Sti} O.Stier, M.Grundmann, and D.Bimberg, 
Phys. Rev. B {\bf 59}, 5688,(1999).
\bibitem{bir}G. L. Bir and G. E. Pikus, Symmetry and Strain Induced Effects 
in Semiconductors (Halsted, United Kingdom, 1974); 
L.D. Landau and E. L. Lifshitz, Theory of Elasticity 
(Addison-Wesley Publishing Company Inc, Reading, Massachusetts, USA, 1970).
\bibitem{pryor}C. Pryor, J. Kim, L.W. Wang, A.J. Williamson and A. Zunger, Phys. Rev. B{\bf 183}, 2548(1998).
\bibitem{adachi}S. Adachi, J. Appl. Phys. {\bf 53}, 8775 (1982).
\bibitem{mathieu}H. Mathieu, P. Meroe, E. L. Amerziane, B. Archilla, J. 
Camassel, and G. Poiblaud, Phys. Rev. B{\bf 19}, 2209 (1979); 
S. Adachi and C. Hamaguchi, Phys. Rev. B{\bf 19}, 938 (1979). 
\bibitem{hinckley}J. M. Hinckley and J. Singh, Phys. Rev. B{\bf 42},
3546 (1990).
\bibitem{land}O. Madelung and M. Schulz, Landolt-B\"{o}rstein, New Series
III/22a, p. 87 (1982).
\bibitem{Cardona} A. Blacha, H. Presting, M Cardona, Phys. Stat. Sol. (b) {\bf
126}, 11 (1984).
\bibitem{Walle} C. G. Van de Walle, Phys. Rev. B {\bf 39}, 1871 (1989).
\bibitem{Eaz} M. Razeghi, Ph. Maurel, F. Omnes, and J. Nagle, Appl. Phys.
Lett. {\bf 51}, 2216 (1987).





\end{thebibliography}
\end{document}


\large
\doublespace

\centerline{\bf \Large 2. MODELING OF STRAIN AND SILO WIRES}

\section{\bf \large   Strain In General. }
\mbox{}

When two lattice-mismatched materials grow on top of each other, strain 
is produced. 

In essence, as long as the lattice
mismatch is below the critical thickness, it is possible to grow an
epitaxial film which is completely in registry with the substrate, i.e., it
is pseudomorphic up to the critical thickness. Beyond the critical thickness,
dislocation are produced and the strain relaxes. It is important to note that
the pseudomorphic films below critical thickness are thermodynamically
stable, i.e., even though they have strain energy, the relaxed state with
dislocations has a higher free energy. This allows one to have large
built-in strain without serious concern that the film will somehow
``disintegrate'' after some time. In this section we will establish 
some basic notions and formulas for strain in crystalline materials. Without losing
any generality, let us consider small values of strain. 

In order
 to  define the strain quantitatively, let us consider an 
 position vector $\vec{r}$
in a Cartesian coordinate system for an atom  in the solid. 
There will be a small displacement $\vec{u(r)}$ when a 
deformation force is applied
on the solid, ie, the new position of the atom is at $\vec{r'}=\vec{r}+
\vec{u}(r)$.
The small change in length element  is
$$dl'^2=d(\vec{r}+\vec{u}(r))^2$$
and let $du_i(r)= \frac{\partial u_i(r)}{\partial x_j} dx_j$, then 
 $$dl'^2=d(\vec{r}+\vec{u(r)})^2= dl^2 + \epsilon_{ij} \cdot dx_idx_j$$
where $\epsilon_{ij}=\frac{1}{2}(\frac{\partial u_i}{\partial x_j}+
\frac{\partial u_j}{\partial x_i}+\frac{\partial u_k}{\partial x_i}\frac
{\partial u_k}{\partial x_j})$ is the strain tensor. 
In small strain, the second order term
is ignored, we have $\epsilon_{ij}=\frac{1}{2}(\frac{\partial u_i}{\partial x_j
}+\frac{\partial u_j}{\partial x_i})$[1].

For our purpose, for  three arbitrary independent vectors which connect atoms 
1,2,3,4 at four corners of a tetragon surrounding an atom(which is placed at
the center), we set up the position matrix for the undeformed tetragon as:
\[
R^0  = \left ( \begin{array}{ccc}
       R^0_{12x} & R^0_{23x} & R^0_{34x} \\
       R^0_{12y} & R^0_{23y} & R^0_{34y} \\
       R^0_{12z} & R^0_{23z} & R^0_{34z}
      \end{array} \right ), \]
 with  ${\bf R}^0_{ij} = {\bf R}^0_j - {\bf R}^0_i; i,j=1,4$ and  and ${\bf R}^0_i$
$(i=1,4)$ denote positions of the four atoms in the system we are considering. 
After applying the deformation force, the new position matrix $R'$ is 
 related to the undeformed $R^0$ as 
$$\vec{R}'= (1+\epsilon)\cdot \vec{R}_0$$ 
with $\epsilon$ being the strain tensor defined for the center atom as 
\[
\epsilon  = \left ( \begin{array}{ccc}
       \epsilon_{xx} & \epsilon_{xy} & \epsilon_{xz} \\
       \epsilon_{yx} & \epsilon_{yy} & \epsilon_{yz} \\
       \epsilon_{zx} & \epsilon_{zy} & \epsilon_{zz}
      \end{array} \right ), \]
which is calculated to be[3] 
\begin{equation}
\epsilon=R'*R_0^{-1} - 1.
\end{equation}
This equation will be used in Chapter 4 for the calculation of strain for
each atoms.

After getting the strain, the effects of the strain on band structure
 and optical properties are
through the deformation potential given by Bir and Pikus[2]. For the effects
on the valance bands, i.e, $p_x$,$p_y$, and $p_z$ in our III-V alloys, we have[2,4]
\begin{equation}
H^{st} = \left( \begin{array}{ccc}
 -\Delta V_H + D_1 &\sqrt{3}d e_{xy} &\sqrt{3}d e_{xz} \\
  \sqrt{3}d e_{xy} & -\Delta V_H + D_2     &\sqrt{3}d e_{yz} \\
       \sqrt{3}d e_{xz} &\sqrt{3}d e_{yz} & -\Delta V_H + D_3
               \end{array} \right),
\end{equation}
where $e_{ij}=(\epsilon_{ij}+\epsilon_{ji})/2$, and
$$
\Delta V_H = (a_1+a_2)(\epsilon_{xx}+\epsilon_{yy}+\epsilon_{zz}),~~\\
 D_1 = b(2\epsilon_{xx}-\epsilon_{yy}-\epsilon_{zz}),~~\\$$
$$ D_2 = b(2\epsilon_{yy}-\epsilon_{xx}-\epsilon_{zz}),~~\\
 D_3 = b(2\epsilon_{zz}-\epsilon_{yy}-\epsilon_{xx}).\\
$$
The strain potential on the $s$ states is given by
$$\Delta V_c = c_1(\epsilon_{xx}+\epsilon_{yy}+\epsilon_{zz}).$$
In general, tensile strain tends to narrow the gap, and push the light hole
band 
up over the heavy hole band; compressive strain leads to the opposite
results.
 
\vspace{2ex}
\section{\bf \large VFF Model of The Growth of SILO Wires. }
\mbox{}

The influence of the strain on the growth of the overlayers are calculated by the
 valence-force-field(VFF) model of Martin[5] and Keating[6]. This model has been 
shown to be successful in fitting and predicting the elastic constants of elastic continuum theory, calculating strain distribution in a quantum well, and determining the
atomic structure of $III-V$ alloys.  It was used in the calculation of the strain
distribution in self-assembled dots[7,8]. The VFF model is a microscopic
theory which includes bond stretching and bond bending, and avoids the potential
failure of elastic continuum theory in the atomistic thin limit. In 
this model, 
the total energy of the lattice is taken as
\begin{equation}
V =\frac{1}{4}\sum_{ij}\frac{3}{4}\alpha_{ij}(d^2_{ij}-d^2_{0,ij})^2/d^2_{0,ij}
+\frac{1}{4}\sum_i\sum_{j\neq k}\frac{3}{4}\beta_{ijk}(\vec{d_{ij}} \cdot
\vec{d_{ik}}
+d_{0,ij}d_{0,ik}/3)^2/d_{0,ij} d_{0,ik},
\end{equation}
\noindent where $i$ runs over all the atomic sites, $j, k$ run over the
nearest-neighbor sites of $i$, $\vec{d_{ij}}$ is the vector joining the
sites $i$ and $j$, $d_{ij}$ is the length of the bond, $d_{0,ij}$ is the
corresponding equilibrium length in the
binary constituents, and $\alpha_{ij}$ and $\beta_{ijk}$ are the
bond stretching and bond bending constants, respectively.
$\alpha$ and $\beta$ are taken from Martin's calculations[5].
For the bond-bending
parameter $\beta$ of In-As-Ga, we take $\beta_{ijk}=\sqrt{\beta[ij]\beta[ik]}
$ following Ref. [5,8].

To find the equilibrium atomic positions  
in the InAs/GaAs self-assembled QWR, we start from
ideal atomic positions and minimize the system energy
using the
Hamiltonian given above. Minimization of the total energy requires one to
solve a set of coupled equations with $3N$ variables, where $N$ is the total
number of atoms. Direct solution of these equations is impractical
in our case, since the system contains more than 6,000 atoms.
We use an approach taken by several authors[4,7,8] which has been shown
to be quite efficient. In the beginning of the simulation all the atoms are
placed on the InP lattice, we allow atoms to deviate from this starting
positions and use periodic boundary conditions in the plane
perpendicular to the growth direction, while keeping atoms in the
planes outside the SPS region at their ideal atomic positions
for an InP lattice (since the self-assembled QWR is grown epitaxially on the InP
substrate). In each iteration, only one atomic
position is displaced and other atomic positions are held fixed.
The direction of the displacement of atom $i$ is determined according to
the force  $f_i=-\partial V/\partial x_i$ acting on it.
All atoms are displaced in sequence. The whole sequence is
repeated until the forces acting on all atoms become zero,
at which point the system energy is a minimum. In this way, 
the equilibrium atomic positions  are simulated.

Once the positions of all the atoms are known, the strain distribution
is obtained through the strain tensor calculated according to the method
described in  Section 2.1.

\vspace{2ex}
\section{\bf \large Empirical Bond-Orbital Model(EBOM). }
\mbox{}

The method used in this thesis for calculating the strained quantum well band
structure is based
on the effective bond-orbital model (EBOM)[10]. 
The effective bond-orbital model employed here is a tight-binding-like model
which includes
nearest-neighbor interactions among bond orbital residing on an fcc lattice.
The version of the
EBOM used in this calculations describes the coupling between the upper four
spin-3/2 valence
bands and the lowest two s-like conduction bands. 
A bond orbital is defined
to be the proper linear combination of the two atomic orbitals within a
zinc-blende crystal which best describes the
states near the center of the Brillouin zone. The parameters that appear in
the theory are given
by a correspondence which is made by requiring that the Hamiltonian in the
bond-orbital basis,
when written for the bulk material and expanded to the second order in
${\bf k}$, agree with the
Luttinger-Kohn expression. For interactions across heterojunctions, we take
the arithmetic average of the matrix elements for the two bulk materials.
Since the nearest-neighbor EBOM parameters for the two materials
are similar, the arithmetic average is a good approximation.
We have also studied the dependence of QWR energies on different
choices of parameters (e.g. geometric average) and found
essentially the same results[11].
EBOM has been successfully applied to the
calculation of electronic states of superlattices[10], strained-layer
superlattices [12,13], semiconductor
quantum wires[14] and dots[15], and impurity levels in quantum dots[15]
and quantum wells[15].
Here, we apply it to this multi-axially strained QWR structures.

  We solve the QWR electronic states according to a procedure similar to
that described in [16].
We first calculate the band structure of a superlattice(SL) with M bilayers of
$Ga_{x}In_{1-x}As$ (the well material) and N bilayers of
$Al_{0.24}Ga_{0.24}In_{0.52}As$
(the barrier material) using the slab method. 
The SL eigenstates with $k_z =0$ can be treated as eigenstates for
an isolated quantum well (QW) and they are denoted
by $|k_1, \bar k_2, n>$, i.e,
\begin{equation}
   H^{QW}|k_1,\bar k_2,n>= E_n(k_1,\bar k_2)|k_1,\bar k_2,n>
\end{equation}
where $H^{QW}$ is the Hamiltonian for the
$Ga_{y}In_{1-y}As$/$Al_{0.24}Ga_{0.24}In_{0.52}As$ QW 
,where $k_1$ and $k_2$ are the two components of the QW in-plane wave vector
along and
perpendicular to the QWR axis, respectively. $k_1$ is a good quantum number
for the QWR array,
while $\bar k_2$ is not. The QWR wave function is a linear combination of SL wave
functions with
the same $k_1$ but with values of $\bar k_2$ differing by 2$\pi/L_2$.
We define $k_2$  to be
the wave number of the QWR array
in the [110] direction, and we can write $\bar k_2= k_2 + 2\pi\zeta/L_2$
with $\zeta=0,1,2,...,$m,
where $m = L_2/a_2$ with $a_2= a/\sqrt{2}$.
The QWR Hamiltonian is $H^{QWR}=H^{SL} + H'$, where $H'= H_{Ga_{x}In_{1-x}As}-
H_{Ga_{y}In_{1-y}As}$ is treated as a perturbation. 
The Hamiltonian matrix for QWR
\begin{equation}
M(n,\zeta|n', \zeta') =
<n',k_1,k_2+ 2\pi\zeta'/L_2|H^{QWR}|k_1,k_2+ 2\pi\zeta/L_2,n>
\end{equation}
\noindent is then diagonalized in the basis consisting of QW eigenstates.

The effect of lateral intermixing of Ga and In is accounted for by using
the virtual-crystal approximation.
Bond-orbital parameters for the inter-diffused materials are
obtained by linearly interpolating
between the values of the parameters for the pure materials.
The optical parameters in these
expressions are determined by requiring that the optical
matrix elements between bulk states
obtained by EBOM be identical to those obtained in the ${\bf k\cdot p}$
theory up to second
order in ${\bf k}$.

 The effect of strain is included by adding a strain Hamiltonian
$H^{st}$ to the EBOM effective Hamiltonian[12,16,17].

\mbox{}

\mbox{}


\large
\doublespace

\centerline{\bf \Large 3. BAND STRUCTURES AND OPTICAL PROPERTIES 
     OF SILO QUANTUM WIRES IN VIRTUAL-CRYSTAL APPROXIMATION }

\section{Introduction}
\mbox{}

$Ga_xIn_{1-x}As$ is one of the most important ternary III-V 
compound semiconductors. Its bandgap covers both the 1.3 and 1.55 $\mu m$ 
range, which are the preferred wavelengths in long distance fiber 
communications[1,2]. However, long-wavelength photonic devices based on 
lattice-matched $Ga_{0.47}In_{0,53}$As/InP heterostructures suffer 
from strong Auger recombination and intervalence band absorption 
processes[7-10].
Recently, to improve the performance of long-wavelength semiconductor 
lasers, long wavelength($\sim 1.55 {\mu}$m) $Ga_xIn_{1-x}As$ quantum-wire 
(QWR) lasers have been grown by a single step molecular beam 
epitaxy technique[1]. It is found that the QWR laser structures are a 
promising choice because of the many predicted benefits, such as higher gain, 
reduced temperature sensitivity, higher modulation bandwidths, and narrower 
spectral linewidths[2]. Short-period superlattices (SPS)  in the direction
perpendicular to the growth direction can be formed  
via the strain-induced lateral-layer ordering 
(SILO) process[1,2]. A quantum wire heterostructure can then be 
created by simply utilizing this SPS structure as the quantum well region in a 
conventional quantum heterostructure. Besides the self-assembled lateral 
ordering, it is believed that the strain also plays a key role[2] 
in the temperature stability and high anisotropy for the QWR laser structure. 
Recent studies indicate that the use 
of the strained-layer quantum wire heterostructure has advantages of 
high-quality interfaces and band-gap tuning which are 
independent of the lattice constant of the 
constituent materials[3,4,6]. Furthermore, much work has been undertaken which 
predicts that by using strained-layer superlattice to form the active 
region of a 
quantum-wire laser, the threshold current can be decreased by one order of 
magnitude, and the optical loss due to intervalence-band absorption and  Auger 
recombination will also be greatly reduced[1,2,7-11]. Also, the temperature 
sensitivity 
is reduced by an order of magnitude compared with strain-free structures[2]. 
A typical 
temperature sensitivity of the lasing wavelength is($\sim 5 \AA/{}^0C$) for 
the usual 
GaInAs/InP lasers. By using a distributed-feedback structure, the temperature 
dependence of the lasing wavelength is reduced to $1\AA/{}^0C$[2]. 
With this strained 
GaInAs QWR, the dependence is smaller than $0.1\AA/{}^0C$[2]. It should be an 
important improvement on the current technology in fabricating the 
long wavelength lasers for fiber communication. 
In this chapter we study the effects of multi-axial strain on
the electronic and optical properties of the QWR structures grown via 
the SILO process. Through this study, we can gain a better 
understanding of the strain engineering of  
QWR structures suitable for the application of fiber-optical communication. 

   In the experiment performed by  Chou et al[1], the QWR active 
region is created 
$\it{in~ situ~}$ by the SILO process within the 
$(GaAs)_2/(InAs)_{2.2}$ SPS regions. 
The SILO process generates a strong Ga/In lateral 
composition modulation and creates In-rich $Ga_xIn_{1-x}As$ lateral QWs in the [110] 
direction. By sandwiching the In composition modulated layer (with $x$ varying 
from 0.3 to 0.7) 
between $Al_{0.24}Ga_{0.24}In_{0.52}As$ barrier layers, a strained QWR 
heterostructure is formed[1,2]. The model QWR structure considered in the 
present paper is depicted in Figure.1. 
The central region consists of the In composition modulated $Ga_xIn_{1-x}As$ 
layer (A/B/A/B...), 
in which the A strip is Ga-rich while the B strip is In-rich. The lateral 
period is $L_2=L_2^A+L_2^B$.
In our calculation, we consider $L_2^A=L_2^B = 150\AA$,
$L_{\perp}^B=60\AA$, and $L_{\perp}^W=100\AA$.
We consider two cases of composition modulation. In case one, the 
Ga composition changes abruptly
with $x=0.6$ (or $0.7$) in strip A and $x=0.4$ (or $0.3$) in strip B. 
In case two, $x$ varies 
sinusoidally from 0.6 (or 0.7) at the 
center of strip A to 0.4 (or 0.3) at the center of strip B 
(with $x=0.5$ at the border between A and B strips).

\section{Theoretical Approach}
\mbox{}

 The method used in this paper for calculating the strained quantum well band 
structure is based 
on the effective bond-orbital model (EBOM)[5] as described in Chapter 2. 
A detailed description of this 
method without 
the effect of strain has been published elsewhere[5]. 

The version of the 
EBOM used in this calculations describes the coupling between the upper four 
spin-3/2 valence 
bands and the lowest two s-like conduction bands. For our QWR system, the 
split-off bands are ignored since the spin-orbital splitting is large. 
The split-off band is about 350 meV below the valence-band maximum.
The coupling energy between the heavy-hole  band and 
split-off band via strain is about 16.47 meV  in our 
system. Based on the second-order perturbation theory, we estimate that the 
inclusion of
split-off band will shift the heavy-hole subbands up by 
approximately 0.8 meV[8].
The effects on the spacings between valence subbands should be much less,
since the energy shifts are similar for the first few valence subbands.

In the diagonalization of the SILO quantum wire Hamiltonian as in Eq.2.5,
the QW basis is truncated
to reduce the computer time, thus making the method a variational one. 
The subbands 
closest to the band edge are converged to within 0.1 meV. The size of the 
Hamiltonian matrix, originally given by the number of sites in a QWR 
supercell 
times the number of coupled bands, is now reduced to the number of QW states 
necessary to give accurate results for the 
QWR of concern. In our case, this number needed to get well converged results 
for the
two uppermost pair of valence subbands is around 200. However, we have used 
$\sim 1200$ basis states 
in order to ensure the accuracy of the deeper lying states for all 
calculations 
in this chapter. 
The effect of lateral intermixing of Ga and In is accounted for by using 
the virtual-crystal approximation.
Bond-orbital parameters for the inter-diffused materials are 
obtained by linearly interpolating 
between the values of the parameters for the pure materials. 
The optical parameters in these 
expressions are determined by requiring that the optical 
matrix elements between bulk states 
obtained by EBOM be identical to those obtained in the ${\bf k\cdot p}$ 
theory up to second 
order in ${\bf k}$[14].

    The effect of strain is included by adding a strain Hamiltonian 
$H^{st}$ to the EBOM effective
Hamiltonian[6,11,16].  The matrix elements 
of $H^{st}$ in the bond-orbital basis can be 
obtained by the deformation-potential theory of Bir and Pikus[17].
Here, we consider the combination
of two bi-axial strains, one in the (001) plane 
(due to the lattice mismatch at the interfaces
between GaInAs and AlGaInAs) and the other in the (110) plane 
(due to the Ga composition modulation ). 
The resulting strains in the QW layers consist 
of both hydrostatic and uniaxial components. For case 1 (Ga composition
changes abruptly from region A to region B), the lattice constant 
of the strained layers in the [001] direction  
($\tilde a_{001}$) is given by minimizing the free energy of 
the fcc system due to strain[8,17]
\begin{eqnarray}
F= \frac{1}{2}C_{11}^A(\epsilon_{11A}^2 + \epsilon_{22A}^2+\epsilon_{33A}^2)
L_2^A~~~~~~~~~~~~~~~~~~~\\\nonumber
~~~~+ C^A_{12}(\epsilon_{11A}\epsilon_{22A}+\epsilon_{22A}\epsilon_{33A}
+\epsilon_{11A}\epsilon_{33A})L_2^A~~~\\\nonumber
+\frac{1}{2}C^B_{11}(\epsilon_{11B}^2 + \epsilon_{22B}^2+\epsilon_{33B}^2)
L_2^B~~~~~~~~~~~~~~~~~\\\nonumber
+ C^B_{12}(\epsilon_{11B}\epsilon_{22B}+\epsilon_{22B}\epsilon_{33B}
+\epsilon_{11B}\epsilon_{33B})L_2^B~~,
\end{eqnarray}
where $L_2^A$ ($L_2^B$) is the lateral layer thickness and $\epsilon_{ijA} 
(\epsilon_{ijB})$ is the strain tensor in the
Ga-rich (In-rich) region. $C^A$'s ($C^B$'s) are the elastic 
constants for the Ga-rich (In-rich) materials.
Here we have used a rotated Cartesian coordinates in
which $x'=[1\bar 1 0], y'=[110],$ and $z'=[001]$.
A constraint $\tilde{a}^A_{\|}=\tilde{a}^B_{\|}=a_{InP}=a_{AlInGaAs}$ 
has been imposed to keep the  $Al_{0.24}Ga_{0.24}In_{0.52}As$ barrier 
strain free, since it is lattice-matched to the InP substrate[1,2],
where $\tilde{a}^\alpha_{\|}$ is the in-plane strained lattice constant 
(perpendicular to [001] axis)
for material $\alpha$. The multi-axial strain caused by the lattice-mismatch
in both (001) and (110) planes is simply
\begin{equation}
\epsilon_{11\alpha} = \epsilon_{22\alpha} =
\frac{\tilde{a}^\alpha_{\|}-a_\alpha}{a_\alpha},
\end{equation}
where $a_A$ ($a_B$) is the unstrained lattice constant of $Ga_{x}In_{1-x}As$
($Ga_{1-x}In_{x}As$) ($x=0.6$ for case 1) 
and 
\begin{equation}
\epsilon_{33\alpha} = \frac{\tilde{a}_{001}-a_\alpha}{a_\alpha}
\end{equation}
with $\epsilon^\alpha_{ij} = 0$ for $i \ne j$.

The minimization procedure leads to 
\begin{equation}
\tilde{a}_{001}=-\frac{1}{2}\frac{ C_{11}^A L_2^A/a_A^2 + 
C_{11}^B L_2^B/a_B^2}{(C_{12}^A \epsilon_{\|}^A L_2^A/a_A+C_{12}^B 
\epsilon_{\|}^B L_2^B/a_B)+(C_{11}^A L_2^A/a_A+C_{11}^B L_2^B/a_B)},
\end{equation}
which is the strained lattice constant in the [001] direction.

Due to symmetry, the multi-axial strain tensor is diagonal 
in the $x'y'z'$ coordinates, which leads to a diagonal
strain Hamiltonian[8] as shown in Chapter 2:
\begin{equation}
H^{st} = \left( \begin{array}{ccc} 
                                        -\Delta V_H + D_1 & 0 & 0 \\
                                          0 & -\Delta V_H + D_2  & 0\\
                                          0 & 0 & -\Delta V_H + D_3
               \end{array} \right),
\end{equation}
where 
$$
\Delta V_H = (a_1+a_2)(\epsilon_{11}+\epsilon_{22}+\epsilon_{33}),~~\\
 D_1 = b(2\epsilon_{11}-\epsilon_{22}-\epsilon_{33}),~~\\$$
$$
 D_2 = b(2\epsilon_{22}-\epsilon_{11}-\epsilon_{33}),~~\\
 D_3 = b(2\epsilon_{33}-\epsilon_{22}-\epsilon_{11})\\
$$
The strain potential on the s states is given by[6]
$$\Delta V_c = c_1(\epsilon_{11}+\epsilon_{22}+\epsilon_{33}),$$
The strain Hamiltonian in the bond-orbital basis $|JM>$ can be easily 
found by using the coupling constants[5], i.e, 
\begin{equation}
<JM|H^{st}|J'M'>=\sum_{\alpha, \alpha',\sigma}C(\alpha, \sigma;J,M)^*
C(\alpha', \sigma;J',M')H^{st}_{\alpha\alpha'}
\end{equation}
The elastic constants $C_{12}$ and $C_{11}$ for GaAs, InAs and AlAs can be found
in Ref. [18,19]. The deformation potentials $a_1,~a_2,~b,~c_1$ can be
found in Ref.[20,21]. The linear interpolation
and virtual crystal approximation is used to obtain the corresponding 
parameters for the GaInAs and AlGaInAs.

	The above strain Hamiltonian is derived for case one only, where
the Ga composition ($x$) changes abruptly from region A to region B. 
For case two, $x$ varies continuously from $x_m$ to $1-x_m$ ($x_m=0.6$ or $0.7$).
In this case, we shall 
first calculate the strain Hamiltonian for the abrupt case 
to get $H^{st}_{x_m}$ and $H^{st}_{1-x_m}$, using Eq. 3.6.
We then obtain the strain Hamiltonian at any atomic layer with Ga composition
$x$ via a linear interpolation between $H^{st}_{x_m}$ and $H^{st}_{1-x_m}$.
This is consistent in spirit with the virtual crystal approximation we have 
used for obtaining the interaction parameters in EBOM. 

\section{Results and Discussions}
\mbox{}

In Figure 3.2, we show
a schematic diagram indicating the alignment between band edges 
of the constituent materials for QWR in case 1 (square profile)
with and without the effect of strain[8]. 
All energies are measured with respect to the bulk InAs valence band edge. 
This diagram is useful in understanding the 
quantum confinement effect on the QWRs considered here.
From Figure 3.2, we notice that the strain [in particular, the one in the (110)
plane] has a significant effect on the band alignment. Take Figure 3.2(a) 
for example,
without strain (solid
lines) the conduction (valence) band of Ga$_{0.4}$In$_{0.6}$As is below (above)
that of Ga$_{0.6}$In$_{0.4}$As by 202 meV (14 meV). 
With strain (dotted lines), the situation is reversed and
the conduction (valence) band of Ga$_{0.4}$In$_{0.6}$As is above (below)
that of Ga$_{0.6}$In$_{0.4}$As by 15 meV (94 meV); thus, 
both electrons and
holes will be confined in the Ga-rich region of the QWR. 

The strain also causes a splitting between the heavy-hole (HH)  and light-hole (LH)
bands with the LH band lying above the HH band. However, the splitting is rather
small, about 3 meV in Ga$_{0.6}$In$_{0.4}$As. When the confinement effect
due to the barrier material (Al$_{0.24}$Ga$_{0.24}$In$_{0.52}$As) is included,
the HH band again lies above the LH band, due to the difference in the effective
masses along the growth direction ($z$). The dash-dotted lines in Figure 3.2(a) 
indicate the superlattice band edges of 100\AA Ga$_{0.6}$In$_{0.4}$As 
(or Ga$_{0.4}$In$_{0.6}$As)
sandwiched between 60\AA {} Al$_{0.24}$Ga$_{0.24}$In$_{0.52}$As barriers. 
The difference in superlattice band edges (dash-dotted lines) between 
Ga$_{0.6}$In$_{0.4}$As and Ga$_{0.4}$In$_{0.6}$As determines the degree of
lateral quantum confinement in the QWR. For the present case, the conduction-band
offset is 20 meV  and the valence-band (for HH only) offset is 
92 meV, as
far as the lateral confinement is concerned. Both offsets are large enough
to give rise to lateral confinement for electrons and holes in
the Ga-rich region. 

Figure 3.3 shows the near zone-center valence subband structures of 
the $100~~~~~~\AA$
Ga$_{x}$In$_{1-x}$As/Al$_{0.24}$Ga$_{0.24}$In$_{0.52}$As quantum well 
with (a) $x=0.6$ and (b) $x=0.7$. The multi-axial strain on 
Ga$_{x}$In$_{1-x}$As is included as if it is in the QWR geometry. 
All subbands are two-fold degenerate due to the Kramer's degeneracy and
they are labeled according to the characters of their underlying Bloch
functions at the zone-center: HH for heavy hole and LH for light-hole.
The valence band structures shown here provide a useful
guideline for understanding
the QWR valence band structures. The black circles indicated energies
at discrete $k_2$ points ($k_2=n 2 \pi/L_2$) which constitute the
zero-th order energies for QWR at the zone center. Due to the combined effect
of strain (which brings HH1 and LH1 subbands close in energy) and the
band-mixing, the HH1 subband has  weak dispersion for $k_2 < 0.02
\AA^{-1}$.  This will cause the top two valence
subbands for the quantum wire to be extremely close in energy.

  Figure 3.4 shows the near zone-center valence subband structures of 
square-shaped QWRs with the multi-axial strain with (a) Ga composition ($x$) 
changing abruptly from 0.6 in region A to 0.4 in region B and (b)Ga composition 
($x$) changing abruptly from 0.7 in region A to 0.3 in region B. 
To compare with experiment[1,2], we 
choose material parameters appropriate for temperature at 77K[13]. 
Similar to the quantum well case, all subbands here are two-fold degenerate 
due to the Kramer's degeneracy.
The conduction subbands are approximately 
parabolic as usual with a zone-center subband minimum equal to
845 meV in (a) and 839 meV in (b) (not shown in the figure). 
This gives an energy gap 847 meV for the (x=0.6/0.4)
QWR array and 797 meV for the (x=0.7/0.3) QWR array . 

The observed C1-HH1 excitonic transition is at 735 meV (or 1.65$\mu m$)
for the QWR[1,2] at 77K. 
In comparing the band gaps with the experiment, one should also take into 
account the exciton binding energy which is around
20 meV for this size of QWR.
Thus, the theoretical result for the (x=0.7/0.3) QWR array is in closer 
agreement with experiment, but still about 40 meV too high. 

   Comparing the band structures in both $k_1$ ($[1\bar 10]$) and $k_2$ 
($[110]$) 
directions, we noticed an apparent anisotropy in the energy dispersion.
The dispersions in the $k_2$ direction for the (confined) valence
bands are rather small, indicating strong lateral confinement.
We observe strong anti-crossing effect between the HH1/HH2 and LH1 subband at
$k_1$ near $0.02 \AA^{-1}$ similar to what happens in a quantum well
[24,25]. 
Here the HH2 subband of the QWR corresponds to the [110] zone-folded  
part of the 
HH1 subband of the quantum well ( the envelope function
has odd parity in the [110] direction but even parity 
in the $[1\bar 10]$ direction). The separation in energy between the
HH1 and HH2 subbands near the zone center is extremely small (about 2 meV)
in this case. This is because the HH1 subband of the quantum well
has almost no dispersion (i.e. with very large effective mass)
in the $k_2$ direction for $k_2 < 0.02 \AA^{-1}$ as a result the
HH1-LH1 band mixing (see Figure 3.3).

	The subbands in Figure 3.4(b) have less dispersion in the $k_2$ 
direction
compared with those in Figure 3.4(a) as a result of stronger lateral 
confinement.
This is caused by the larger band discontinuities in the x=0.3/0.7 case (versus
the x=0.4/0.6 case) as can be seen by comparing band alignments shown in 
Figure 3.2.

     Figure 3.5 shows the near zone-center valence subband structures of 
a QWR with a sinusoidal lateral modulation with Ga composition ($x$)
varying as the position ($y'$) in the [110] direction between two extreme values
$x_m$ and $1-x_m$, i.e. $$x(y')=0.5 + (0.5-x_m) sin(2\pi y'/L_2)$$ for
$x_m=0.4$ and (b) $x_m=0.3$.  
In this case, the Hamiltonian including the multi-axial strain is 
calculated with a linear extroplation between the Hamiltonians
for the maximum and minimum Ga composition, i.e.
$$
H_{Ga_{x}In_{1-x}As}^{st}=f H_{Ga_{x_m}In_{1-x_m}As} + 
(1-f) H_{Ga_{1-x_m}In_{x_m}As},
$$
where $x_m=0.4$ or 0.3 and $f$ is determined by comparing the Ga composition 
on both sides of the equation
$$ 
x = x_m f + (1-x_m) (1-f),
$$
or $f= (1-x_m-x)/(1-2x_m)$.

The conduction band minimum (not shown) is 846 meV (842 meV) for the
case $x_m=0.4$ ($x_m=0.3$), which corresponds to
a band gap of 853 meV (807 meV). Subtracting the exciton binding energy 
($\sim$ 20 meV)
from the band gap, we found that the $x_m=0.3$ case is in closer agreement 
with the observed excitonic transition energy of 735 meV.
The band gaps of QWR's with the sinusoidal profile are consistently larger
(by about 10 meV)
than the corresponding QWR's with the square-profile (with the same extreme 
values of Ga composition). 
The sinusoidal variation in Ga composition gives rise to more energy, 
since the well region contains more materials with band gap higher 
than the minimum in comparison with the square-profile.
Furthermore, the lateral strain is reduced (less mismatch on the average) 
which also tend to increase the band gap.

	Comparing Figure 3.5 with Figure 3.4, we notice that the spacing
between HH1 and HH2 is substantially larger for the sinusoidal profile
than for the square profile. This can be understood by the following argument.
The envelope function for the HH2 state is more spread out than that for
the HH1 state, thus its energy is increased more in the sinusoidal profile
(with higher probability being in materials with higher band gap)
compared with the square profile.

	Next we discuss the optical properties of QWR's. Since the QWR states
can be qualitatively viewed as the zone-folded states of the quantum well (QW) 
states (the zeroth-order states) via the lateral confinement, it is instructive 
to examine the optical matrix elements of the QW case. Figure 3.6 shows the
squared optical matrix elements of the 100\AA
Ga$_{0.7}$In$_{0.3}$As/Al$_{0.24}$Ga$_{0.24}$In$_{0.52}$As quantum well
for the HH1-C1 and LH1-C1 transitions versus the wave vector $k_2$. 
The solid and dashed lines are for
the polarization vector along the $k_1$ ($[1\bar 1 0]$) and $k_2$ ([110])
directions, respectively. We note that the optical matrix elements are isotropic
in the $x$-$y$ plane at the zone-center, while they become anisotropic at
finite values of $k_2$. At the zone center, the HH state consists of  
Bloch states with atomic character $(x'+iy')\uparrow$ ($|J,M>=|3/2, 3/2>$), 
while the LH state  consists of Bloch states with atomic character 
$(x'-iy')\uparrow+2z\downarrow$ ($|J,M>=|3/2,-1/2>$); 
thus, the corresponding optical transitions 
(to an s-like conduction band state) are isotropic in the $x$-$y$ plane.
Here $x' (y')$ is the coordinate along the $k_1 (k_2)$ direction.
For finite $k_2$, the HH and LH characters are mixed, with the HH state 
consisting of more $x'$-character than $y'$-character, thus in favor of
the polarization vector parallel to the $k_1$ direction. Note that the HH (LH) 
band tends to have an atomic character with polarization perpendicular 
(parallel) to the direction of wave vector. This is a direct consequence
of the fact that the ($pp\pi$) interaction is weaker that the ($pp\sigma$) 
interaction in a tight-binding model.
Since QWR states are derived from the QW states with
finite values of $k_2$, we expect the optical matrix elements of QWR
to be anisotropic in the $x$-$y$ plane as well.

Figures 3.7 and 3.8 show the squared optical matrix elements versus wave vectors 
of the corresponding QWR's considered in Figures 3.4 and 3.5, respectively.
respectively. The solid and dashed lines are for
the polarization vector along the $k_1$ ($[1\bar 1 0]$) (parallel to
the wire) and $k_2$ ([110]) (perpendicular to the wire) directions.  
The optical matrix elements together with the subband structures  discussed 
above provide the essential ingredients for understanding the optical 
transitions observed in the photoluminescence (PL) measurements. 
In these figures the optical transitions considered are
from the topmost three valence subbands to the lowest conduction subband: 
HH1-C1, HH2-C1, and LH1-C1. These curves were obtained by summing 
the contributions of two degenerate subbands (due to Kramer's degeneracy)
for the initial and final states.
To compare with experimental (PL) results, we concentrate on the HH1-C1 
transition. 

From Figures 3.7 and 3.8,  we found for the parallel polarization (solid lines), 
the squared optical matrix element for the HH1-C1 transition has a maximum at 
the zone center with a value near 10 eV and remains close to this
value for all finite $k_2$ and with $k_1=0$. For the
perpendicular polarization, the HH1-C1 transition is very small for
finite $k_2$ and $k_1=0$.  This means that the wave function of the HH1 state 
has mostly $x'$ character (and some $z$ character). 
Thus, we conclude that the strong 
lateral confinement forces the wave function of the HH1 state with $k_1=0$ to 
change from the $x'+iy'$ character (in the QW case) into mostly $x'$ character.
At finite $k_1$, the character of the HH1 state gradually changes back
toward  the $x'+iy'$ character and the optical transition becomes almost 
isotropic when $k_1$ is comparable to zone-boundary value of 
$k_2 (\pi/L_2)$, and finally turns into mostly $y'$ character as
$k_1$ becomes much larger than $(\pi/L_2)$. The
anti-crossing behavior of the HH1/HH2 subbands with the LH1 subband 
further complicates the whole picture at $k_1 \approx 2 \AA^{-1}$.

The HH2-C1 transition has zero optical strength at the zone center for
both polarizations. This is expected, since the HH2 state has odd parity in the
envelope function, which leads to forbidden transition at the zone center.
The symmetry restriction is relaxed as the wave vector deviate from zero.
The LH1-C1 transition has large optical strength (around 17 eV)
for the perpendicular
polarization and weak strength (around 4 eV) for the parallel 
polarization at the zone center.
This indicates that the LH1 states consists of mostly $y'$ character.
So the strong lateral confinement forces the the wave function of the LH1 state 
with $k_1=0$ to change from the $x'-iy'$ character (in the QW case) 
into mostly $y'$ character. In other words, the HH1 state (with pure 
$x'+iy'$ character) and LH1 state (with pure $x'-iy'$ character) in the QW case
are mixed thoroughly by the lateral confinement in QWR to produce
a predominantly $x'$ state (HH1) and a predominantly $y'$ (LH1) state.
Note that all these states consist of appreciable $z$ character, which
will appear in the optical transition with $z$ polarization.

To calculate the anisotropy, we integrate the squared optical matrix 
element over range of $k_1$ corresponding to the spread of exciton 
envelope function in the $k_1$ space and average over two $k_2$
points. Because the lateral confinement for the quantum wire is
strong, the optical anisotropy was found to be rather insensitive
to the variation in $k_2$. Thus, the two point average in $k_2$ 
space is adequate. The exciton envelope function
is obtained by solving the 1D Schr\"{o}ding equation for the exciton in
the effective-mass approximation
\begin{equation} 
[ - \frac {\hbar^2} {2\mu} (\frac  {\partial} {\partial x'})^2
 + V_X(x')]F(x')=E_X F(x'), \end{equation}
where $\mu$ is the exciton reduced mass ($\approx 0.049 m_0$)
and $V_X(x')$ is the effective 1D exciton potential given by[24]
\[ V_X(x')=\frac {e^2}{\epsilon_0 |x'|}(1-e^{-\beta|x'|}). \]
The parameter $\beta$ is obtained by extrapolation from the values given in
Table V of Ref. \onlinecite{Sanders1}. For a $150\AA$ quantum wire, we obtain 
$\beta=0.03 \AA^{-1}$. Eq. (10) is solved numerically, and the exciton
envelope function in $k_1$ space is obtained via the Fourier transform of
$F(x')$. The exciton binding energy obtained is 17 meV.
We found that the ratio of the averaged  
optical strength for the HH1-C1 transition for the parallel to perpendicular 
component of the polarization vector is 3.41 (5.48) for the $x=0.4/0.6$ 
($x=0.3/0.7$)
QWR with square profile, and 5.60 (12.3) for the $x_m=0.4$ ($x_m=0.3$)
QWR with sinusoidal profile.  The $x_m=0.3$ case shows stronger
optical anisotropy than the $x_m=0.4$ case, indicating that the stronger 
lateral confinement leads to stronger optical anisotropy as expected.
Experimentally, the optical anisotropy is found to be around 2-4. 
So, our results are consistent with experiment. 

\section{Conclusion}
\mbox{}

We have calculated the band structures and optical matrix elements 
for the strained QWR grown by the SILO method. The effect of multi-axial strain 
on the valence subbands and optical matrix elements is discussed.  
Our theoretical studies provide the explanation of the anisotropy in optical 
matrix elements of these QWR's observed experimentally.
We find that the biaxial strain due to the lattice mismatch between the
Ga-rich and In-rich regions is most dominant. It 
tends to increase the lateral confinement in the QWR and enhances the 
anisotropy of the optical transitions which may be useful 
for certain applications in optical communication. We
also calculated the effect of lateral composition modulation on the band 
structures and optical properties and find that it increases the band gaps
and  the optical anisotropy. 


\mbox{}
\mbox{}

\newpage

\begin{figure}[b]
\caption{\bf \large Schematic sketch of the QWR array fabricated in Refs. [1,2]. 
The QWR axis lies in the $[1\bar 10]$ direction. A and B label 
the Ga-rich and In-rich strips, respectively.}
\end{figure} 

\mbox{}
\mbox{} 

\clearpage

\begin{figure}[b]
\begin{small}

\caption{\bf \large Schematic diagram 
indicating the alignment between band edges 
of the constiuent materials for QWR in case 1 (square profile)
with and without the effect of strain. 
Solid lines are for unstrained bulk,
dotted lines are for bulk under multi-axial strain appropriate for the
present QWR, and dash-dotted lines are for a 100\AA 
GaInAs/AlGaInAs quantum well  including the multi-axial strain. 
}
\end{small}
\end{figure}

\mbox{} 
\mbox{} 

\clearpage
\begin{figure}[b]
\caption{\bf \large Valance subband structures for the 100\AA
Ga$_{x}$In$_{1-x}$As/Al$_{0.24}$Ga$_{0.24}$In$_{0.52}$As quantum well  
with (a) $x=0.6$ and (b) $x=0.7$.
}
\end{figure}

\mbox{} 
\mbox{} 

\newpage

\begin{figure}[b]
\caption{\bf \large Valance subband structures for QWR's with square profile
for (a) Ga-composition $x=0.6$ in Ga-rich and $x=0.4$ in In-rich region and
(b) $x=0.7$ in Ga-rich and $x=0.3$ in In-rich region.
}
\end{figure}

\mbox{} 
\mbox{} 

\newpage
\begin{figure}[b]
\caption{\bf \large  Valance subband structures for QWR's with 
sinusoidal profile with $x$ ranging from $x_m$ to $1-x_m$ 
for (a) $x_m=0.6$ and (b) $x_m=0.7$.
}
\end{figure}

\mbox{} 
\mbox{} 

\newpage
\begin{figure}[b]
\caption{\bf \large Squared optical matrix elements for transitions from HH1 
and LH1 to the first conduction subband of the 100\AA
Ga$_{0.7}$In$_{0.3}$As/Al$_{0.24}$Ga$_{0.24}$In$_{0.52}$As quantum well  
for light polarized parallel (solid), perpendicular (dashed) 
to the QWR axis and z component (dash-dotted).            
}
\end{figure}

\mbox{} 
\mbox{} 

\newpage
\begin{figure}[b]
\caption{\bf \large
 Squared optical matrix elements for transitions from 
the top three valence subbands to the first conduction subband
for light polarized parallel (solid) and perpendicular (dashed) 
to the QWR axis for QWR's considered in Fig. 3.3.
}
\end{figure}

\mbox{} 
\mbox{} 

\newpage
\begin{figure}[b]
\caption{\bf \large
 Squared optical matrix elements for transitions from 
the top three valence subbands to the first conduction subband
for light polarized parallel (solid) and perpendicular (dashed) 
to the QWR axis for QWR's considered in Fig. 3.4.
}
\end{figure}


\large
\doublespace

\centerline{\bf \Large 5. INTER-BAND AND INTER-SUBBAND OPTICAL TRANSITIONS IN
SELF-ASSEMBLED QUANTUM WIRES}

\section{Introduction}
\mbox{}

	Quantum-well infra-red photodetectors (QWIP's) have been
extensively studied in recent years.\cite{Lev}${}^-$\cite{kapon} The main mechanism used in
QWIPs is the inter-subband optical transition, because the wavelengths
for these transitions in typical III-V quantum wells can be tailored
to match the desired operating wavelength 
for infra-red (IR) detection. Due to its narrow band absorption,
QWIP's are complementary to the traditional HgCdTe detectors, which
utilize the inter-band absorption, and therefore are applicable 
only for broad-band absorption.
QWIP's based on III-V semiconductor quantum wells
have been shown to be promising candidate
for making flat-panel high-density IR imaging arrays used in 
IR detectors.\cite{Liu}
The main drawback of QWIP's is the lack of normal-incidence
capability, unless some processing is made to create 
diffraction gratings on the surface, which tends to reduce the responsivity
of the material to the incident radiation.\cite{Liu} 
Because electrons in
quantum wells have translational invariance (within the effective-mass model)
in the plane normal to the growth axis, the electron
inter-subband transitions for normal-incident radiation is zero 
(or very small even if the coupling with other bands is considered).
One way to break the translational invariance is to introduce the
surface diffraction grating as commonly adopted in many QWIP's fabricated today.
A better (and less expensive) way to break the in-plane translational
invariance is to utilize the strain-induced lateral modulation provided in 
self-assembled nano-structure materials. These nano-structures include
quantum wires\cite{ky1}$^-$\cite{wohlert1} and quantum dots.
\cite{Eag}$^-$\cite{Sti}
Because the lateral modulation is formed
via self-assembly, the fabrication of this type of materials will be much 
more efficient once the optimized growth parameters are known. Hence, it will be
cost effective to use them for device fabrications.

Self-assembled III-V QWR's grown via the strain-induced lateral-layer ordering
(SILO) process have attracted a great deal of attention recently.
\cite{chou}${}^-$\cite{wohlert1}
The self-assembly process occurs during the growth of
short-period superlattices (SPS) [e.g. (GaAs)$_2$/(InAs)$_{2.25}$]
along the [001] direction on an InP substrate. 
The excess fractional InAs layer leads to stripe-like islands 
during the initial MBE growth.\cite{chou} The presence of stripes combined with strain
leads to natural phase separation as
additional layers of GaAs or InAs are deposited and the structure becomes
laterally modulated in terms of In/Ga composition.
A self-assembled QWR heterostructure can then be
created by sandwiching the laterally modulated layer
between barrier materials such as Al$_{0.24}$Ga$_{0.24}$In$_{0.52}$As 
(quarternary), Al$_{0.48}$In$_{0.52}$As (ternary), or InP
(binary).\cite{chou}${}^-$\cite{wohlert1}

It was found that different barrier materials can lead to different degree of
lateral composition modulation, and the period of lateral
modulation ranges from 100\AA \, to 300 \AA \, depending on
the growth time and temperature.\cite{KY}

	In this chapter, we explore the usefulness 
of InGaAs quantum wires (QWR's) grown by the strain-induced lateral ordering 
(SILO) process for IR detection. 
Our theoretical modeling includes the effects of realistic band 
structures and microscopic strain distributions by combining the effective
bond-orbital model (EBOM)\cite{Chang} 
with the valence-force-field (VFF) model.\cite{VFF1}${}^-$\cite{VFF3}
One of the major parameters for the IR detectors is the absorption 
quantum efficiency which is directly related to the absorption
coefficient by $\eta=1- e^{-\alpha l}$ where $\alpha$ is the absorption
coefficient and $l$ is the sample length.\cite{IR} Thus, to have a realistic
assessment of the material for device application, we need to
perform detailed calculations of the absorption coefficient, 
taking into account the excitonic and band structure effects. 
Both inter-subband and inter-band transitions
are examined systematically for a number of structure parameters 
(within the experimentally feasible range) chosen
to give the desired effect for IR detection. 

	It is found that the wavelengths for the
inter-subband transitions of InGaAs self-assembled QWR's range from 
10 to 20 $\mu m$,
while the inter-band transitions are less than 1.5 $\mu m$. Thus, the material
provides simultaneous IR detection at two contrasting wavelengths,
something desirable for application in multi-colored IR video camera.

Several structure models with varying degrees of alloy mixing for lateral 
modulation are considered. For the inter-band absorption, the excitonic
effect is important, since it gives rise to a large shift in transition energy
and substantial enhancement of
the absorption spectrum.
To study the excitonic effect on the absorption spectrum for both discrete and
continuum states,  we use a large set of basis functions with a finite-mesh
sampling in the k-space and diagonalize the exciton Hamiltonian directly.
Emphasis is put on the analysis of line shapes of various peak structures
arising from discrete excitonic states of one pair of subbands coupled
with the excitonic (discrete and continuum) states associated with other
pairs of subbands. We find that the excitonic effect enhances 
the first absorption peak around 1.2-1.5 times and shifts
the peak position by 20-50meV. 

\section {Theoretical model}
\mbox{}

The QWR structures considered here consist of 8 pairs of
(GaAs)$_2$(InAs)$_{2.25}$ short-period superlattices (SPS) sandwiched between
Al$_{0.24}$Ga$_{0.24}$In$_{0.52}$As  barriers. 
The SPS structure prior to  strain induced lateral
ordering (SILO) is depicted in Fig. 4.1(b).
With lateral ordering, the structure is 
modeled by a periodic modulation of alloy composition
in layers with fractional monolayer of (In or Ga) in the SPS structure.
In layers 7 and 9 (starting from the bottom as layer 1), we have   
\begin{equation}
x_{\mbox{In}}=  \left\{ \begin{array}{ll}
 x_m [1-\sin (\pi y'/2b)] /2 &  \mbox{ for } y'< b\\ 
	0  & \mbox{ for } b < y' < L/2-b \\
 x_m \{1 + \sin [\pi (y'-L/2)/2b]\} /2 &  \mbox{ for } L/2-b < y' < L/2+b \\
 x_m   &	  \mbox{ for } L/2+b < y' < L-b \\
 x_m \{1 - \sin [\pi (y'-L)/2b]\} /2 &  \mbox{ for } y'> L-b, \end{array} \right.
\end{equation}
where $x_m$ is the maximum In composition in the layer, $2b$ denotes the 
width of lateral composition grading, and $L$ is the period of the
lateral modulation in the [110] direction. The experimental feasible range of
$L$ is between 100 \AA and 300 \AA. The length of $L$ is controlled by
the growth time and temperature.

In layers 3 and 13, we have
\begin{equation}
x_{\mbox{In}}=  \left\{ \begin{array}{ll}
 0   &    \mbox{ for } 0 < y' < 5L/8-b \\
 x_m \{1 + \sin [\pi (y'-5L/8)/2b]\} /2 &  \mbox{ for } 5L/8-b<y'< 5L/8+b,\\ 
 	x_m  & \mbox{ for } 5L/8+b < y' < 7L/8-b \\
 x_m \{1 - \sin [\pi (y'-7L/8)/2b]\} /2 &  \mbox{ for } 7L/8-b < y' < 7L/8+b \\
 0   &    \mbox{ for } 7L/8+b < y' < L.
\end{array} \right.
\end{equation}
Similar equation for $x_{\mbox{Ga}}$ in layers 5 and 11 can be deduced
from the above.
By varying the parameters $x_m$ and $b$, we can get different degrees
of lateral alloy mixing. Typically $x_m$ is between 0.6 and 1, and $b$
is between zero and $15 a_{[110]} \approx$ 62 \AA.

A VFF model\cite{VFF1}${}^-$\cite{VFF3} is used to find the equilibrium 
atomic positions in the self-assembled QWR structure by minimizing the lattice energy.
The strain tensor at each atomic (In or Ga) site is then obtained by
calculating the local distortion of chemical bonds.
Once the microscopic strain distribution in the model structure is determined,
the energy levels and wave-functions of self-assembled 
quantum wires are then calculated within the effective bond-orbital model 
(EBOM). Detailed description of this method can be found in Refs. 
\cite{Chang,Houng,Cit,Sun}. 
EBOM used here is a tight-binding-like model in which two s-
like conduction bands (including spin) and four valence bands 
with total angular momentum $J=3/2$ (due to spin-orbit coupling
of p-like orbitals  with the spinor).  Thus, 
the present model is comparable to the six-band ${\bf k\cdot p}$
model.

To minimize the computing effort, we express 
the electron and hole states for the quantum wire
structures in terms of eigen-states of a quantum well structure
with different in-plane wave vectors. 
The quantum well consists of 8 pairs of (GaAs)$_2$(InAs)$_2$ short-period
superlattice (SPS) plus two InAs monolayers (one inserted after the second
pair of SPS and the other after the sixth pair of SPS), so the the total
In/Ga composition ratio is consistent with the (GaAs)$_2$(InAs)$_{2.25}$ SPS.
The whole stack of SPS's is then sandwiched
between two slabs of Al$_{0.24}$Ga$_{0.24}$In$_{0.52}$ barriers.  
Let us denote the quantum well eigen-states 
as $|n,k_1,k_2\rangle_{QW}$ where $n$ labels the subbands, 
$k_1$ denotes the wave vector
along the wire ($[1 \bar 1 0]$) direction 
and $k_2$ labels the wave vector in the [110]
direction, which is perpendicular to the
wire and the growth axis. Expanding the quantum well states in terms of
bond-orbitals, we have
\[ |n,k_1,k_2\rangle_{QW}=\frac 1 {\sqrt{L_s}} \sum_{\alpha,{\bf R}}
f_{n,k_1,k_2}(\alpha,R_z)\exp(ik_2 R_2)
\exp(ik_1 R_1)|u_{\alpha}({\bf R})\rangle, \]
where $L_s$ is the sample length along the wire axis, 
$f_{n,k_1,k_2}(\alpha,R_z)$ is the eigen-vector for the quantum well
Hamiltonian, and $u_{\alpha}({\bf R})$ denotes
an $\alpha$-like bond orbital state at site ${\bf R}$ ($\alpha=1,\cdots,6$
for two s-like conduction-band and four $J=3/2$ 
valence-band orbitals). Here ${\bf R}$ runs over all lattice sites within
the SPS layer (well region) and AlGaInAs  layer (barrier region).

We  then diagonalize
the hamiltonian for the quantum wire (QWR) within a basis which consists of
the quantum well states with $k_2$'s separated by reciprocal lattice vectors
$g_m = m (2\pi/a_{[110]})$; $m$= integers. Thus, the QWR states are written 
as
\[ |i,k_1,k_2\rangle= \sum_{n,m} C_{i,k_1}(n,k_2+g_m)
|n,k_1,k_2+g_m\rangle_{QW} \]
where $C_{i,k_1}(n,k_2+g_m)$ is the eigen-vector for the 
quantum-wire hamiltonian
matrix for the $i-th$ QWR subband at wave vector $(k_1,k_2)$.
 
In terms of the bond orbitals, we can rewrite the QWR states as

\[ |i,k_1,k_2 \rangle=\sum_{\alpha,{\bf R}} F_{i,k_1,k_2}
(\alpha,{\bf R}) |u_{\alpha}({\bf R})\rangle \]
where
\[ F_{i,k_1,k_2}(\alpha,{\bf R})=\frac 1 {\sqrt{L}} 
\sum_{n,m} C_{i,k_1}(n,k_2+g_m)f_{n,k_1,k_2+g_m}(\alpha,R_z)
\exp[i(k_2+g_m)R_2] \exp(ik_1 R_1) \]
is the QWR envelope function.
For the laterally confined states, the dispersion of bands versus $k_2$
is negligible; thus, the $k_2$ dependence can be ignored.

The absorption coefficient for inter-subband transitions between 
subbands $i$ and $j$ is given by
\begin{small}
\begin{equation}
 \alpha_{ij}(\hbar\omega)=\frac{4\pi^2e^2\hbar}{n_r~ m_e^2~c~L_z~\hbar\omega}
\sum_{k_1,k_2}|\langle i,k_1,k_2|\hat{\epsilon}
\cdot {\bf p}|j,k_1,k_2\rangle|^2
[f_i(k_1,k_2)-f_j(k_1,k_2)] \delta(E_j(k_1,k_2)-E_i(k_1,k_2)-\hbar\omega) 
\end{equation}
\end{small}
where $n_r$ is the refractive index of the 
QWR, $L_z$ is the width of the SPS region (or QW confined in the $z$
direction), $f_i(f_j)$ is the Fermi-Dirac distribution function for
subband $i$ ($j$). The optical matrix elements between QWR subband states
are related to those between bond orbitals by 
\[ \langle i,k_1,k_2|\hat{\epsilon} \cdot {\bf p}|j,k_1,k_2\rangle = 
\sum_{\alpha,\alpha',\tau} F^*_{i,k_1,k_2}(\alpha,{\bf R}) 
F_{j,k_1,k_2}(\alpha',{\bf R})
\langle u_{\alpha}({\bf R})|\hat{\epsilon}\cdot {\bf p}|
u_{\alpha'}({\bf R}+\vec \tau) \rangle, \]
where $\vec \tau$ runs over on-site or the 12 nearest-neighbor sites 
in the fcc lattice.
The optical matrix elements between bond orbitals are related to the 
band parameters by requiring the optical matrix elements between bulk
states near the zone center to be identical to those obtained in
the ${\bf k \cdot p}$ theory\cite{kdp}. We obtain\cite{Cit}
\[ \langle u_{s}({\bf R})|p_{\alpha}
u_{\alpha'}({\bf R}) \rangle = \sqrt{m_0 E_p/2}
\delta_{\alpha,\alpha'}; \alpha=x,y,z, \]
\[ \langle u_{s}({\bf R})|p_{\alpha}
u_{s}({\bf R}+\vec \tau) \rangle = (\hbar/2\sqrt{2}a)(E_p/Eg-m_0/m_e^*) 
\tau_{\alpha} ; \alpha=x,y,z, \]
where $\tau_{\alpha}$ is the $\alpha$-th of the lattice vector $\tau$
in units of a/2, $E_p$ is the inter-band optical matrix element as defined in
Ref. \onlinecite{kdp}, and $m_e^*$ is the electron effective mass.

Next, we study the inter-band transitions. For this case, the excitonic
effect is important for undoped samples. For heavily doped samples, 
the excitonic
effect can be ignored due to the phase-spacing filling and carrier
screening effects. We shall consider both situations.
Here we are only interested in the absorption spectrum
near the band edge due to laterally confined states. Thus, the dispersion
in the $k_2$ direction can be ignored. Without the excitonic effects, the 
absorption coefficient is given by Eq. (5.3) with $i$ ($j$) indicating
the valence (conduction) subband states. At low temperatures, the valence
subbands are completely filled, so the Fermi-Dirac distribution ($f_i$)
in Eq. (5.3) can be replaced by 1. 
For undoped samples, the exciton
states with zero center-of-mass momentum
is written as linear combinations of products of
electron and holes states associated with the same $k_1$
(wave vector along the wire direction). We write the QWR electron-hole
product state for the $i$-th conduction
subband and $j$-th valence subband as

\[ |i,j;k_1\rangle =|i,k_1\rangle |j,k_1\rangle\]
\[\equiv \sum_{\alpha,\beta,{\bf R}_e,{\bf R}_h}F_{i,k_1}(\alpha,{\bf R}_e)G_{j,k_1}
(\beta,{\bf R}_h)|u(\alpha,{\bf R}_e)>|u(\beta,{\bf R}_h)>.\]
               
The matrix elements of the exciton Hamiltonian within this basis is given by
\begin{small}
\begin{equation}
   \langle i,j;k_1|H_{ex}|i',j';k_1'\rangle = 
[E_i(k_1)-E_j(k_1)]\delta_{i,i'}\delta_{j,j'}\delta_{k_1,k'_1}
- \sum_{{\bf R}_e,{\bf R}_h} {\cal F}_{ii'}^*(k_1,k'_1;{\bf R}_e)v({\bf R}_e
-{\bf R}_h){\cal G}_{jj'}(k_1,k'_1;{\bf R}_h),\end{equation}
\end{small}
where $v({\bf R}_e,{\bf R}_h)=\frac{4\pi~e^2}{\epsilon(0)~|{\bf R}_e-
{\bf R}_h|}$ is
the coulomb interaction between the electron and hole screened by
the static dielectric constant $\epsilon(0)$, and
\[ {\cal F}_{ii'}(k_1,k'_1;{\bf R}_e) = \sum_{\alpha}F^*_{i,k_1}(\alpha,{\bf R}_e)
F_{i',k'_1}(\alpha,{\bf R}_e) \]
describes the charge density matrix for the electrons.
Similarly, 
\[ {\cal G}_{jj'}(k_1,k'_1;{\bf R}_h) = \sum_{\beta}G^*_{j,k_1}(\beta,{\bf R}_h)
G_{j',k'_1}(\beta,{\bf R}_h) \]
describes the charge density matrix for the holes.
In Eq. (5.4), we have adopted the approximation
\[   \langle u(\alpha,{\bf R}_e), u(\beta,{\bf R}_h)|v
|u(\alpha',{\bf R}'_e),u(\beta',{\bf R}'_h)\rangle             
\approx v({\bf R}_e-{\bf R}_h)\delta_{\alpha,\alpha'}\delta_{\beta,\beta'}
\delta_{{\bf R}_e,{\bf R}'_e} \delta_{{\bf R}_h,{\bf R}'_h}, \]
since the Coulomb potential is a smooth function over the distance of 
a lattice constant, except at the origin, and the bond orbitals are
orthonormal to each other.
At the origin (${\bf R}_e={\bf R}_h$), the potential is singular, and
we replace it by an empirical constant which is adjusted so as to give
the same exciton binding energy
as obtained in the effective-mass theory for a bulk
system. The results are actually insensitive to the on-site Coulomb
potential parameter, since the Bohr radius of the exciton is much larger
than the lattice constant.

After the diagonalization, we obtain the excitonic states as linear 
combinations of the electron-hole product states, and the 
inter-band absorption coefficient is computed according to

\begin{equation}
\alpha^{ex}(\hbar\omega)=\frac{4\pi^2e^2\hbar}{n_r~ m_e^2~c~A~\hbar\omega}
\sum_{n}|\sum_{k_1,i,j}C^n_{ij}(k_1)\hat{\epsilon}\cdot {\bf p}_{ij}(k_1)|^2\delta(
\hbar\omega-E^{ex}_{n}),
\end{equation}
where $A$ is the cross-sectional area of the SPS region within
the QWR unit cell (as depicted in Fig. 5.1). $C^n_{ij}(k_1)$ is the
$n$-th eigen-vector obtained by diagonalizing the exciton Hamiltonian of Eq.
(5.4). The momentum matrix elements ${\bf p}_{ij}(k_1)$ are given by 

\begin{equation}
{\bf p}_{ij}(k_1)=
\sum_{\alpha,\beta,\vec \tau}F^*_{i,k_1}(\alpha,{\bf R}_e) 
G_{j,k_1}(\beta,{\bf R}_h+\vec \tau)
\langle u_{\alpha}({\bf R}_e)|\hat{\epsilon}\cdot {\bf p}|
u_{\beta}({\bf R}_h+\vec \tau) \rangle. \end{equation}
In the absence of band mixing, the conduction subband state reduces to a pure
$s$-like state and the valence subband state reduces to a pure $p$-like state.
In that case, only the element $\langle u_{s}({\bf
R}_e)|p_{\alpha}|u_{p_{\alpha}}({\bf R}_e) \rangle = \sqrt{m_0E_p/2} $ 
is needed, and the results become similar to that obtained by
the effective mass method.

In order to obtain a smooth absorption spectrum, we replace the $\delta$ 
function in Eq. (5.1)
by a Lorentzian function with a half-width $\Gamma$, 

\begin{equation}
\delta(E_i-E)\approx \Gamma/\{\pi[(E_i-E)^2+\Gamma^2]\},
\end{equation}
where $\Gamma$ is energy width due to inhomogeneous broadening, which is taken to be 
0.02 eV.

\section {Results And Discussions}
\mbox{}

       We have performed calculations of inter-subband and inter-band 
absorption spectra[36] for
the QWR structure depicted in Fig. 4.1(b) with varying degree of alloy mixing and
different lengths of period ($L$) in the lateral modulation. We find that
the inter-subband absorption
spectra are sensitive to the length of period ($L$), but rather insensitive to
the degree of alloy mixing. Thus, we only present results for
the case with moderate alloy mixing, which are characterized by parameters
$b=33\AA$ and $x_m=1.0$. In the calculations of strain distribution
via the VFF model, the bottom layer atoms of QWR's
are bounded by the InP substrate, while the upper layer atoms and GaAs capping 
layer atoms are allowed to move freely. 
This structure corresponds to the
unclamped structure as indicated in Ref. 34.
	
	For different period length ($L$) of QWR's, 
the strain distribution profiles 
are qualitatively similar as shown in Ref. 34. As $L$ decreases, 
the hydrostatic strain in In-rich region (the right half of QWR unit cell
shown in Fig. 4.1(b)) increases, while it decreases in the Ga-rich region (the left
half of QWR unit cell). 
The bi-axial strain components vary with $L$ in the opposite way
in comparison to the hydrostatic component. 

	Figure 5.1 depicts
the potential profiles due to strain-induced lateral ordering
seen by an electron in two  QWR structures considered here ($L=50 a_{[110]}$
and $L=40 a_{[110]}$). Solid(Dashed) lines are for model structures 
with (without) alloy mixing. From the potential profile, we see that both
the electron and hole in the ground state are laterally confined in
the In-rich region (the right half of the QWR unit cell). 

	The conduction subband structures for the self-assembled QWRs with 
alloy mixing
($x_m=1.0$ and $b=8a_{[110]}$) for ($L=50 a_{[110]}$
and $L=40 a_{[110]}$) are shown in Fig. 5.2. All subband are grouped
in pairs with a weak spin splitting (not resolved on the scale shown).
The first (C1) and third (C3) pairs 
of subbands are nearly dispersionless along the $k_2$ direction, 
while the second (C2) and fourth (C4) pairs
of subbands have appreciable dispersion. Comparing the envelope functions
for the first four pairs of conduction subband states (C1-C4)
with the solutions to rectangular quantum box solutions labeled by
principal quantum numbers $(n_y,n_z)$, we can identify them as
the (1,1), (2,1), (1,2), and (3,1) states, respectively.
Thus, the C1 and C3 states are well confined in the lateral direction, 
and the corresponding subbands are nearly dispersionless in $k_2$. 
On the other hand, the  C2 and C4 states are 
less confined in the $y'$ direction due to the larger spread
of their $y'$ component wave function (with higher principal quantum numbers)
and the corresponding subbands have appreciable
dispersion along $k_2$. The inter-subband transition between the first
two pairs (C1-C2) gives rise to the dominant IR response. 
For $L=40 a_{[110]}$, the C2 and C3 subbands are nearly
degenerate for $k_2=0$ (due to accidental degeneracy), and the (2,1) and
(1,2) characters are strongly mixed. However, for finite $k_2$, the
degeneracy is removed [due to the large dispersion  of the (2,1) state] and
they are no longer mixed.

	The valence subband structures for the self-assembled QWRs with 
alloy mixing ($x_m=1.0$ and $b=33 \AA$) for ($L=50 a_{[110]}$
and $L=40 a_{[110]}$) are shown in Fig. 5.3. All valence subbands display
substantial spin splitting along $k_1$ and the first four pairs ($V1-V4$) of
subbands are nearly dispersionless along $k_2$, indicating strong
lateral confinement effect.  The valence subbands with energies below $
-130 meV$ correspond to unconfined states; thus, they are closely spaced.

\vspace{2ex}
\noindent {\bf  A. Inter-subband absorption }
\vspace{2ex}

	The inter-subband absorption
spectrum is the most relevant quantity in determining
the usefulness of self-assembled QWR's for application in IR
detection. Fig. 5.4 shows the calculated inter-subband absorption spectra
of the self-assembled QWR structure (as depicted in Fig. 4.1(b)) for three 
different lengths
of period: $L=72, 50$, and $40 a_{[110]}$ (approximately 300 \AA, 200 \AA, and
160 \AA, respectively). 
In the calculation, we assume that these QWR structures are n-type
doped with linear carrier density around $1.2 \times 10^6cm^{-1}$ 
(which corresponds to
a Fermi level around 25 meV above the conduction band minimum).

For comparison purposes, we show results for
polarization vector along both the [110] (solid curves)
and [001] directions (dashed curves). The results for [1$\bar 1$0]
polarization are zero due to the strict translational invariance imposed in our
model calculation.  

The  peak positions for the
inter-subband transition with normal incidence (for [110] polarization) are 
around 65 meV, 95 meV, and  125 meV for the three cases considered
here. All these are within the desirable range of IR detection. 
As expected, the transition energy increases as the length of 
period decreases due to the increased degree of lateral confinement. 
However, the transition energy will saturate at around 125 meV as we
further reduce the length of period, since the bound-to-continuum
transition is already reached at $L=40 a_{[110]}$.

The absorption strength for the [110]-polarized peak decreases
from 13000 cm$^{-1}$ for $L=72 a_{[110]}$ to 7000 cm$^{-1}$ for 
$L=40 a_{[110]}$. This is mainly due to the fact that $\alpha$ is inversely
proportional to $\hbar \omega$ [see Eq. (5.2)]  and the the peak becomes
more spread out in energy as caused by the increased dispersion in C2
subband for narrower QWR. 
The inter-subband absorption for the [001] polarization is peaked around
125 meV for all three cases.
The excited state involved in this transition is the C3 subband which
correspond to the $(n_y,n_z)=(1,2)$ state .
Thus, it has the same physical origin as the inter-subband
transition used in typical QWIP structure.
Although this peak is not useful for IR detection with normal
incidence, it can be used as the second-color detection
if one puts a diffraction grating on the surface as typically done 
in the fabrication of QWIPs. In this respect, the QWR IR detector
(for cases 1 and 2) is unique
in that it has two strong narrow-band absorption at two
distinct wavelengths, which are also polarization dependent.
For case 3, the energy positions for the (1,2) and (2,1) states nearly
coincide (both are near the top of the quantum well); thus, the
inter-subband absorption spectra for [110] and [001] polarizations are
peaked at almost the same position.
The absorption strength for the [001]-polarized peak increases
from 11000 cm$^{-1}$ for $L=72 a_{[110]}$ to 18000 cm$^{-1}$ for 
$L=40 a_{[110]}$. This is mainly due to the fact that $\alpha$ is 
inversely proportional to the cross-sectional area of the QWR.

\vspace{4ex}
\noindent {\bf B. Inter-band absorption }
\vspace{2ex}

	The inter-band optical transitions are important for the
characterization of self-assembled QWR's, since they are readily observable via
the Photoluminescence (PL) or optical transmission experiment.
For IR-detector application, they offer another absorption peak at
mid IR wavelengths, which can be used together with the inter-subband
transitions occurred at far IR wavelengths for multi-colored detection.
Thus, to understand the full capability of the self-assembled QWR material, we
also need to analyze the inter-band absorption.


	Fig. 5.5 shows the inter-band absorption spectra for SILO
QWR's with (a) $L=72 a_{[110]}$, (b) $L=50 a_{[110]}$, and 
(c) $L=40 a_{[110]}$. 
Both doped and undoped samples are considered. The results
for the undoped samples are denoted by solid (with exciton effect)
and dashed (without the exciton effect) curves. The results for
doped samples (with $1.2 \times 10^6 cm^{-1}$) are indicated by dotted curves. 
The upper group of curves labeled $(x,y)-pol.$ are for absorption averaged over
the $x$ and $y$ polarizations.  The lower group labeled $z-pol.$ are
for polarization along the $z$ direction.
The PL properties of the $L=72 a_{[110]}$ structure with alloy mixing
characterized by $x_m=1.0$ and $b=33 \AA$ has been studied
in our previous paper.\cite{li1} The QWR structure has a gap around 0.74 eV
with a PL polarization ratio ($P_{[1\bar 10]}/P_{[110]}$) around 2.2. 
The absorption coefficient for this structure with excitonic effect has
a peak strength around $10^4$ cm$^{-1}$. The binding energy
for the ground state exciton (derived primarily from
the top valence subband and the lowest
conduction subband) is around 20 meV. Thus, the peak position in the
absorption spectrum shifts from
0.76 meV (without the excitonic effect) to 0.74 meV 
(with the excitonic effect). The excitonic effect also 
enhances the peak strength by about 20\%.

For the QWR structures with $L=50 a_{[110]}$ and $L=40 a_{[110]}$, we obtain
similar absorption spectra with slightly stronger exciton peaks.
The exciton binding is around 30 (40) meV, and the 
excitonic enhancement factor of the first peak is around 1.35 (1.47) for the
$L=50 (40) a_{[110]}$ case. 
The enhanced peak strength and larger exciton binding energy are caused by the 
stronger lateral confinement for electrons and holes,.
Comparing the dashed curves with the dotted curves, we can see the 
effect of phase-space filling due to doping. For samples grown for mid-IR 
detector application with doping levels below $1.2 \times 10^{6}$ cm$^{-1}$, 
the absorption spectrum should vary between the undoped case (solid curve) and
the heavily doped case (dashed curve) presented here. There is no experiments
so far for these structures we designed, and the experiments on these
IR detectors are highly desirable. 

\newpage

\section {Summary}
\mbox{}

We have studied the inter-subband and inter-band
absorption spectra for self-assembled InGaAs quantum wires
for consideration in IR-detector application. 
Detailed band structures, microscopic strain distributions, and excitonic
effects all have been taken into account.
A number of realistic structures grown via strain-induced lateral 
ordering process are
examined. We find that the self-assembled InGaAs quantum wires 
are good candidate for multi-colored IR detector materials.
They offer two groups of strong IR absorption peaks: one in the 
far-IR range with wavelengths covering 10 - 20 $\mu m$ (via
the inter-subband transition), the other in the mid-IR range with
wavelengths shorter than 1.5 $\mu m$ (via the inter-band
transitions). Due to the strain induced lateral modulation, 
the inter-subband transition
is strong for normal incident light with polarization along
the direction of lateral modulation ([110]). 
The absorption strength is around
$10^4$ cm$^{-1}$ for a doping level of 
$1.2 \times 10^{6} cm^{-1}$. 
This  gives the self-assembled InGaAs quantum wires 
a distinct advantage over the
quantum well systems for application in IR detection.

	The inter-subband absorption is found to be sensitive
to the length of period ($L$) of lateral modulation with the absorption
peak position varying from 60 meV to 120 meV as the length of period
is reduced from 300 \AA \, to 160 \AA. However, further reduction
in the length of period does not shift the absorption peak very much, as
the excited states become laterally unconfined.

	For the inter-band transition, we find that the excitonic effect
enhances the absorption peak strength by about 20-50 \%,
and shift the peak position
by about  20-50 meV for the undoped structures considered. These results
are useful for characterization of the material and for the design of 
mid-IR photo-detectors. We find that
the reduction in the period length ($L$) leads to 
stronger lateral confinement, hence larger exciton
binding and stronger absorption strength. For the doped samples whose
inter-subband transitions are studied, the inter-band transition (without
excitonic effect) gives a broad band absorption in the mid-IR range
(with wavelengths shorter than 1.5 $\mu m$) with a strength between 
4000 and 10000 cm$^{-1}$. The simultaneous presence of this mid-IR absorption 
and the narrow band far-IR absorption near 10$\mu m$ provide a unique
combination for multi-color IR detection application.  

In conclusion, we have successfully demonstrated that self-assembled 
quantum wires
are promising IR-detector materials and we have provided 
theoretical modeling for
the optical characteristics of realistic QWR structures, which can be used to
guide future fabrication of quantum wire infrared detectors.

\newpage
\vspace{2cm}


\clearpage
\mbox{}
\vspace{5.0in}
\begin{figure}[b]
\caption{\bf \large Conduction band and valence band edges for
self-assembled QWR structure depicted in Fig. 1
for (a) $L=50 a_{[110]}$ and (b) $L=40 a_{[110]}$. Dashed: without alloy
mixing. Solid: with alloy mixing described by $x_m=1.0$ and $b=8a_{[110]}$.
}
\end{figure}

\clearpage
\mbox{}
\vspace{5.0in}
\begin{figure}[b]
\caption{\bf \large  Conduction subband structure of 
self-assembled QWR for (a) $L=50 a_{[110]}$ and  (b) $L=40 a_{[110]}$
with $x_m=1.0$ and $b=8a_{[110]}$.
}
\end{figure}

\clearpage
\mbox{}
\vspace{5.0in}
\begin{figure}[b]
\caption{\bf \large Valence subband structure of 
self-assembled QWR for (a) $L=50 a_{[110]}$ and  (b) $L=40 a_{[110]}$
with $x_m=1.0$ and $b=8a_{[110]}$.
}
\end{figure}

\clearpage
\mbox{}
\vspace{5.0in}
\begin{figure}[b]
\caption{\bf \large Inter-subband absorption spectra of 
self-assembled QWR for (a) $L=72 a_{[110]}$, (b) $L=50 a_{[110]}$, and  
(c) $L=40 a_{[110]}$ with $x_m=1.0$ and $b=8a_{[110]}$. 
Solid: [110] polarization, dashed : [001] polarization.
}
\end{figure}


\clearpage
\mbox{}
\vspace{5.0in}
\begin{figure}[b]
\caption{\bf \large  Inter-band absorption spectra of
self-assembled QWR's for (a) $L=72 a_{[110]}$, (b) $L=50 a_{[110]}$,
and (c) $L=40 a_{[110]}$ 
with $x_m=1.0$ and $b=8a_{[110]}$. 
Solid:  undped sample with excitonic effect. 
Dashed: undoped sample without excitonic effect. 
Dotted : doped sample without excitonic effect.  
The set of curves with larger strength are for unpolarized light 
with normal incidence. The set of curves with smaller strength are
for $[001]$ polarization.
}
\end{figure}